\documentclass{spie}

\usepackage{amsmath,amsfonts,amssymb}
\usepackage{graphicx}
\usepackage[colorlinks=true, allcolors=blue]{hyperref}
\usepackage{sidecap}
\usepackage[dvipsnames]{xcolor}
\usepackage{multirow}

\def\arcsec{\hbox{$^{\prime\prime}$}}
\def\deg{\hbox{$^\circ$}}

\def\lae{\mathrel{\raise .4ex\hbox{\rlap{$<$}\lower 1.2ex\hbox{$\sim$}}}}
\def\gae{\mathrel{\raise .4ex\hbox{\rlap{$>$}\lower 1.2ex\hbox{$\sim$}}}}

\def\aap{{A\&A,~}}

\def\aj{{AJ,~}}
\def\apj{{ApJ,~}}

\def\ao{{Appl.Optics,~}}
\def\apss{{Ap\&SS,~}}
\def\mnras{{MNRAS,~}}
\def\nat{{Nature,~}}
\def\pasj{{PASJ,~}}

\def\prd{{PhysRevD,~}}
\newcommand\araa{ARA\&A,~}

\def\procspie{{SPIE Conference Series,~}}

\newcommand{\axaf}{\mbox{\em Chandra\/}}

\newcommand{\redsox}{the {\em REDSoX} Polarimeter}
\newcommand{\rs}{{\em REDSoX}}

\newcommand{\sxp}{{\em PiSoX}}
\newcommand{\ixpe}{\mbox{\em IXPE\/}}

\title{A Small Satellite Version of a Broad-band Soft X-ray Polarimeter} 

\author[a]{Herman L.\ Marshall}
\author[a]{Sarah N.\ T.\ Heine}
\author[a]{Alan Garner}
\author[b]{Eric M.\ Gullikson}
\author[a]{H.\ Moritz G\"unther}
\author[c]{Christopher Leitz}
\author[a]{Rebecca Masterson}
\author[a]{Eric D.\ Miller}
\author[d]{William Zhang}
\author[e]{Rozenn Boissay-Malaquin}
\author[f]{Ilaria Caiazzo}
\author[a]{Deepto Chakrabarty}
\author[a]{Rosemary Davidson}
\author[g]{Luigi C. Gallo}
\author[a]{Ralf K. Heilmann}
\author[h]{Jeremy Heyl}
\author[a]{Erin Kara}
\author[j]{Alan Marscher}
\author[a]{Norbert S. Schulz}
\affil[a]{MIT Kavli Institute, Cambridge, MA, USA}
\affil[b]{Lawrence Berkeley National Lab, Berkeley, CA, USA}
\affil[c]{MIT Lincoln Lab, Lexington, MA, USA}
\affil[d]{Goddard Space Flight Center, Greenbelt, MD, USA}
\affil[e]{University of Maryland Baltimore County, Baltimore, MD, USA}
\affil[f]{California Inst. of Technology, Pasadena, CA, USA}
\affil[g]{Saint Mary's University, Halifax, NS, Canada}
\affil[h]{University of British Columbia, Vancouver, BC, Canada}
\affil[j]{Boston University, Boston, MA, USA}

\authorinfo{Further author information: (Send correspondence to H.L.M.)\\H.L.M.: hermanm@space.mit.edu,
Telephone: 1 617 253 8573} 

\begin{document}

\maketitle

\begin{abstract}

We describe a new implementation of a broad-band
soft X-ray polarimeter, substantially based on a previous design.
This implementation, the Pioneer Soft X-ray Polarimeter (\sxp) is a SmallSat,
designed for NASA's call for Astrophysics
Pioneers, small missions that could be CubeSats, balloon experiments, or SmallSats.
As in \rs, the grating arrangement is designed optimally for the purpose of polarimetry
with broad-band focussing optics by matching the dispersion of the spectrometer channels
to laterally graded multilayers (LGMLs).
The system can achieve polarization modulation factors over 90\%.
For \sxp, the optics are lightweight Si mirrors in a one-bounce parabolic configuration.
High efficiency, blazed gratings from
opposite sectors are oriented to disperse to a LGML forming a channel
covering the wavelength range from 35 \AA\ to 75 \AA\ (165 - 350 eV).
Upon satellite rotation, the intensities of the dispersed spectra, after reflection and
polarizing by the LGMLs, give the three Stokes parameters needed to
determine a source's linear polarization fraction and orientation.
The design can be extended to higher energies as LGMLs are developed
further.  We describe examples of the potential scientific return from instruments
based on this design.

\end{abstract}


\label{sec:summary}

Like true pioneers, we can explore a new frontier in astrophysics.
This instrument would be the first spectropolarimeter
for the soft X-ray band, the Pioneer Soft X-ray Polarimeter (\sxp).
Observing neutron stars with \sxp, we
will test the effects of vacuum birefringence,
search for proton cyclotron lines, and check atmospheric models --
ultimately constraining the equation of state of nuclear matter.
Using \sxp\ to observe active galaxies, we will
discriminate between jet and disk models of their very soft X-ray emission,
providing measures of jet magnetic fields, disk inclinations, and
the importance of Comptonization.
\sxp\ will expand our view of polarized emission across the electromagnetic
spectrum, complementing results from the
Imaging X-ray Polarimetry Explorer (\ixpe, 2-8 keV \cite{ixpe}),
by measuring spectral components that cannot be examined with \ixpe.

\sxp\ is a mission made feasible by
our recent technology advancements in
X-ray mirrors, gratings, and multilayer mirrors.
The science instrument in \sxp\ consists of
parabolic X-ray optics
and Critical Angle Transmission (CAT)
gratings that
disperse soft X-rays to a polarizing multilayer (ML) coated
mirror that then reflects the light to a CCD detector.
\sxp\ will be the
first orbital telescope using Si meta-shell optics and the first
instrument to use CAT gratings and laterally graded ML mirrors in space.
These technologies have been matured in the lab to TRL 5-6 as
part of development
for missions like Arcus or Lynx. \sxp\ will provide invaluable
flight experience for these technologies, yet can be executed with minimal risk given the
significant prior investments by NASA.

The dispersion of the gratings is matched
to the lateral grading of the ML mirror
set at 45$\deg$ to the optical axis.
By smoothly rotating the instrument about the optical axis through 180\deg, we
can measure the three Stokes parameters $I$, $Q$, and $U$ across the entire band
and discern wavelengths with a spectral resolution $\lambda/\delta \lambda \approx 100$.

The spacecraft will be provided by NanoAvionics with flight-proven components.
\sxp\ is designed for the ESPA Grande rideshare to
a low Earth orbit with modest inclination.
We have a 1 year observing plan that will satisfy the science goals,
achieving minimum detectable polarizations (MDPs) of 3-10\%
for over a dozen targets.

\sxp\ is a mission to examine neutron star atmospheres and magnetospheres
and the jets and disks of active galaxies -- key 2010 Decadal Survey topics.
\sxp\ leverages several new but highly developed, now low-risk technologies
with a team that advances the readiness of early-career
researchers and students to assume roles in advancing NASA’s strategic objectives.

\section{Science Objectives}

Polarimetry provides two extra dimensions that have been underutilized in astronomy.
If included, it can elucidate the underlying physics and geometry of the
emission region much better than non-polarimetric observations.
Generally, light can be polarized when it is emitted in or passes through areas with strong magnetic fields
or scatters, e.g.\ off electrons in accretion disk atmospheres.
There are many examples of the value of
optical and radio polarimetry such as
finding hidden broad line regions in quasars \cite{1985ApJ...297..621A},
demonstrating that jets and supernova
remnants accelerate electrons to very high energies \cite{1959AJ.....64Q.339M,2019ARA&A..57..467B},
and examining the
emission regions of pulsar magnetospheres \cite{radha}.

\subsection{Magnetized Atmospheres of Neutron Stars (NSs) and Pulsars}

\begin{figure}
    \centering
    \includegraphics[width=8.5cm]{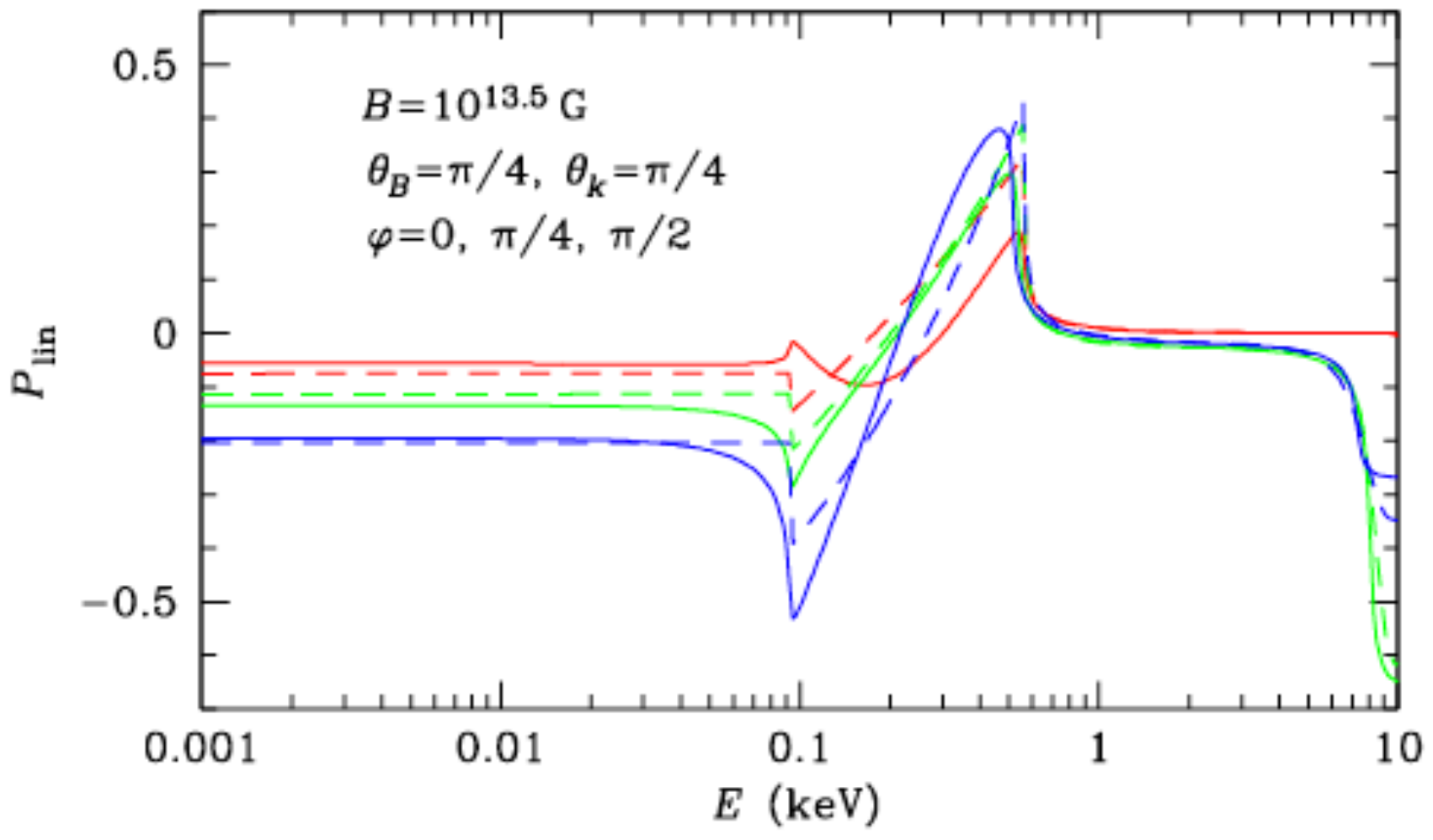}
    \includegraphics[width=7.5cm]{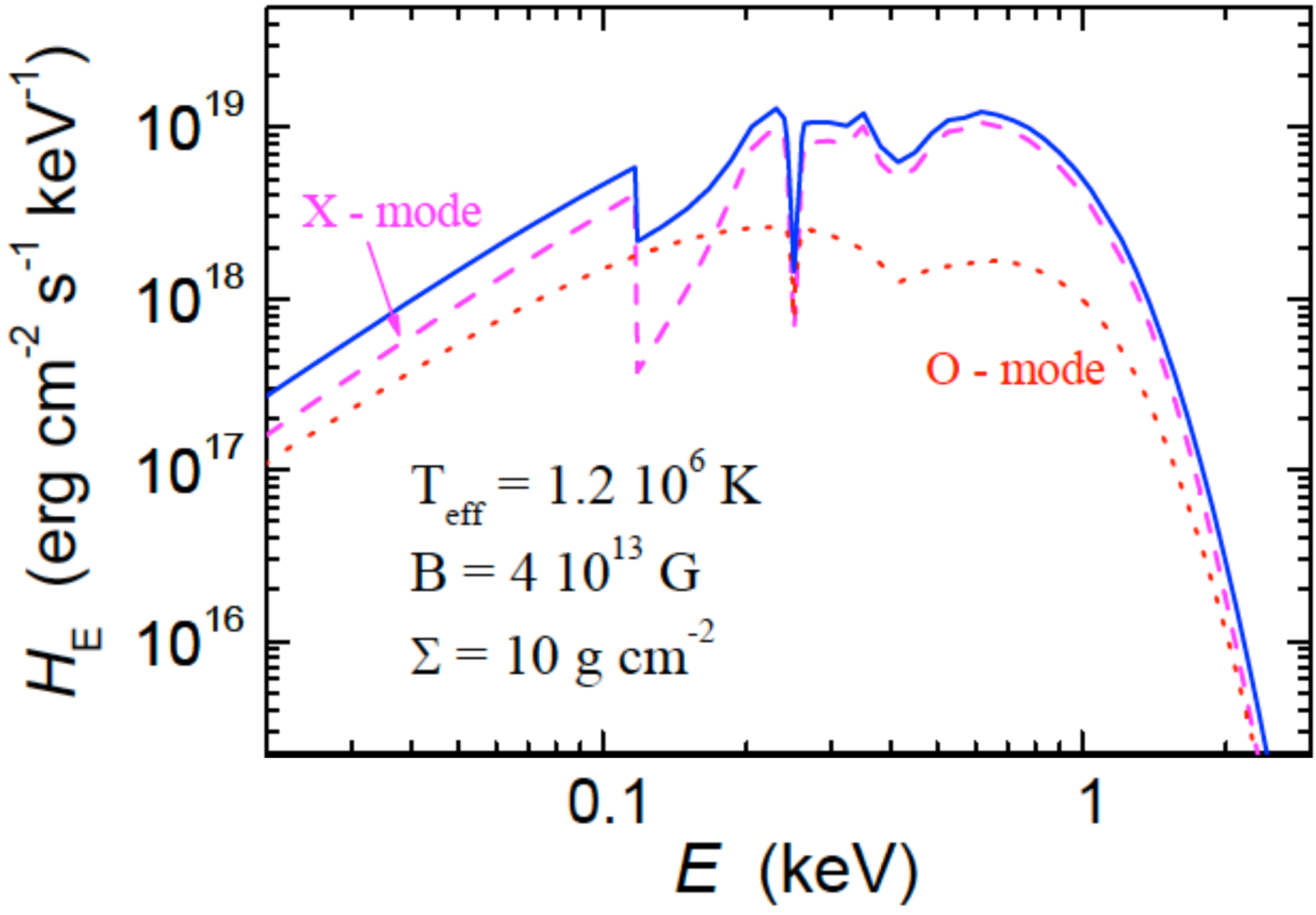}
 \caption{\footnotesize 
{\em Left:} Models of the polarization fraction, given as the
intensity-normalized Stokes $Q$, and its variation with energy
for a magnetized Fe atmosphere with $B = 10^{13.5}$ G
\cite{2012A&A...546A.121P}. (Dashed lines are analytical approximations.)  The values in the legend
represent the angle of the field relative to the surface normal, $\theta_B$, the
polar and azimuthal angles
of photon propagation relative to the surface normal, $\theta_k$ and $\varphi$.
The angle $\varphi$ can be determined from
the soft X-ray polarization variation with $E$, measurable with \sxp.
{\em Right:} Model of an atmosphere showing how the emission lines appear
differently in different polarizations, given by the extraordinary (X) and
ordinary (O) modes of photon propagation relative to the local
magnetic field \cite{2010A&A...522A.111S}.
 }
\label{fig:nsatm}

\end{figure}

Some of the strongest magnetic fields in the universe are observed in isolated NSs, making
them unique laboratories for the interaction of matter and fields, as well as for
testing quantum electrodynamics (QED) effects \cite{2006RPPh...69.2631H}.
\sxp, can be used to sample a large range
of NS $B$-fields with different targets to
understand the origin of atmospheric absorption lines, verify QED predictions,
test atmosphere models, and assess effects in pulsar magnetospheres.
Recently,
atmosphere modeling generally includes linear polarization but this
aspect of the models cannot be tested currently.
These models are needed in order to accurately
determine a NS radius, $R$, that is key to testing equations of state of matter
at nuclear densities 
\cite{2006RPPh...69.2631H,2008AIPC..983..340H,2010A&A...522A.111S}.
\sxp\ will provide the missing polarimetry information
needed to test these atmosphere models (see Fig.~\ref{fig:nsatm}).
\ixpe\ will observe NSs to examine magnetospheric phenomena and the effects of
QED on the pulsed radiation but the NS surfaces
are observable only below 1 keV, which makes these
measurements impossible for \ixpe\ but perfect for \sxp.

Furthermore,
identifying an atmospheric absorption feature by observing
the polarization across the feature would provide
the gravitational redshift at the NS surface, $z = (1-2GM/[Rc^2])^{-1/2} - 1$,
giving the NS mass $M$ when $R$ is determined
from atmosphere fitting.  Measuring $z$ this way has never been done
for any NS (although claimed \cite{2002Natur.420...51C} but not confirmed \cite{2008ApJ...672..504C}),
so \sxp\ provides a ground-breaking opportunity.
Determining both $M$ and $R$ for any NS is critical if we ever hope to use NSs to
study the fundamental nature of matter itself.
As listed in Table~\ref{tab:targets}, several of the candidate targets have
absorption features in their phase-averaged 0.25-0.5 keV spectra
that are thought to arise from proton cyclotron features
when $B \gae 10^{13}$ G \cite{rxj0720,2007ApSS.308..181H,2017MNRAS.468.2975B}
or gravitationally redshifted O VIII \cite{2012MNRAS.419.1525H}.
The centroid of a deep absorption feature in RX J1308.6$+$2127
varied with pulse
phase, perhaps due to a multipolar magnetic field \cite{2017MNRAS.468.2975B}.
As the lines appear to be resolved with FWHM of $\sim 150$ eV level \cite{rxj0720,2017MNRAS.468.2975B},
we require a spectral resolution of better than 50 eV
in order to measure the polarization change through the line, as predicted
in atmosphere models (Fig.~\ref{fig:nsatm} right; \cite{2009A&A...500..891S,2010A&A...522A.111S}).
To resolve these lines requires splitting the spectrum into 4-10 bands,
requiring spectrally integrated MDPs of $\le$ 10\% in order to yield MDPs of $\le 30$\% per band.

Even without absorption features, the polarization fraction
and position angle (PA) vary with energy within the \sxp\ band in a manner
that depends on the atmosphere's composition and whether the atmosphere is
gaseous, liquid or solid (Fig.~\ref{fig:nsatm} left, \cite{2012A&A...546A.121P}).
At these high magnetic fields, the vacuum is birefringent
\cite{2002PhRvD..66b3002H,hsl03}, polarizing the X-rays from the
surface up to 80\% \cite{2015MNRAS.454.3254T} for a gaseous atmosphere and 20-50\% for a condensed surface.  To temporally resolve the
variations requires
MDP $<$ 30\% in 10 pulse phase bins, or a pulse averaged MDP of $\le$ 10\%.
With periods in the 3-20 s range \cite{2019A&A...622A..61S}, we
require X-ray time-tagging to $<$ 0.1 s.

PSR B0656$+$14, and Her X-1 are more ``traditional'' pulsars with $B \sim 10^{12}$ G.
An observation of PSR B0656$+$14 could distinguish between fan and pencil beam
models of magnetospheric emission via rotation of polarization angle $\phi$ with pulse phase and
constrain NS spin axis inclination and $B$-field obliquity \cite{meszaros88}.
Pulsars will be observed by \ixpe\ but the X-ray emission of
PSR B0656$+$14 peaks near 0.3 keV and is too faint above 2 keV for \ixpe.
While most models assume that $B$ is dipolar, rapid changes of the polarization angle
(PA) would indicate that field multipoles are important near the surface.
The sub-keV pulsed emission of Her X-1 does not come from the surface or polar cap, as its effective size is much larger than the NS, and is more likely from the inner edge of the warped,
highly inclined accretion disk where the magnetosphere disrupts it \cite{2005ApJ...633.1064H}.
Two scenarios can only be distinguished using \sxp:
Either the beamed hard X-rays heat a spot on the disk, creating a scattering atmosphere up to 11\% polarized \cite{chandra60} or the sweeping magnetic field heats the ionized disk and the radiation is predominantly synchrotron, up to 70\% polarized if the field is very ordered there.  Either way, the polarization PA would vary significantly with pulse phase and would probe the state and geometry of the transition region where material flows from the disk into the accretion column. Short \sxp\ observations would easily detect polarization even in the case of a disk scattering atmosphere.

\subsection{Soft X-rays from Active Galactic Nuclei (AGN)}

The soft X-ray emission of active galaxies can result from several
different mechanisms that can be distinguished with observations by \sxp,
ranging from strongly magnetized jets to mildly polarized accretion
disk atmospheres and coronae that current observations cannot discriminate.
The soft spectral components of these sources makes \sxp\ an excellent
instrument for exploring models of the soft X-ray emission.

\subsubsection{Relativistic Jets from Blazars}

\label{bllscience}

Blazars, which include BL Lac objects and highly variable quasars,
contain parsec-scale jets with
$\beta \equiv v/c \sim 0.98$ or higher.
Blazar jets are among the most powerful outflows in the universe.
While the jet launching mechanism is not
completely understood, it is clear that magnetic fields must be involved via the
Blandford-Payne or Blandford-Znajek mechanisms \cite{2019ARA&A..57..467B}.
Their X-ray spectra are often quite steep --
power law spectra with spectral energy indices $\alpha = 1.5-2$.
With \sxp, we will measure the uniformity and orientation of the
magnetic field in blazar jets to constrain their origin and evolution.

In the so-called high synchrotron peak blazars (HSPs such as Mk 421),
the X-ray spectrum is most likely synchrotron radiation from high energy electrons.
If the soft X-ray polarization is
much less than $\sim$70\%, then models with simple, uniformly magnetized X-ray emission
regions will be ruled out.
Jet and shock models predict different $B$-field directions and, consequently,
PA.  For knots in a laminar jet flow,
$B$ can be nearly parallel to the jet axis \cite{1980ApJ...235..386M,2004ApJ...600..127G}, while for
shocks $B$ should be perpendicular \cite{1985ApJ...298..114M}.
Both cases were observed in the optical and radio polarimetry of the M 87 jet
\cite{1999AJ....117.2185P,2016ApJ...832....3A},
indicating highly ordered magnetic fields in regions that also produce X-ray emission \cite{2002ApJ...564..683M}.
To allow for low ordering as observed in the UV for some HSPs, we require
MDPs $\le$ 10\%.

In low synchrotron peak blazars (LSPs such as 3C 273), the
emission above 2 keV is
thought to be dominated by inverse Compton (iC) emission or
X-ray reflection from an accretion disk.
Notably, 3C 273 has a separate spectral component
that dominates below 1 keV, whose polarization cannot be examined by \ixpe.
If it is synchrotron emission, related to the superluminal jet in the core,
then the soft X-rays will be more highly
polarized than the hard X-ray emission -- up to 70\%.
If, instead, it is related to the accretion disk, then the polarization
would be below 10\%.
An MDP of 10\% should discriminate these cases.

\subsubsection{AGN Soft Excesses}

\label{sec:agnscience}

Some AGN, such as the subclass known as narrow line Seyfert 1s (NLS1s) \cite{gallo18},
have an X-ray spectral component below 1 keV -- a soft X-ray excess
over a power law component that dominates the spectrum above 2 keV
\cite{1985MNRAS.217..105A}.
The flatter power law that dominates the 2-10 keV band is
most often modeled as reflection of X-rays from a hot
corona off of
an accretion disk \cite{1993MNRAS.261..346H,2012MNRAS.426L.101M},
and its polarization properties are observable by \ixpe.
The polarizations of soft excesses, however, would only be
measurable by \sxp.

There has been a significant debate in the past few decades regarding the
nature of these soft excesses.
They have been variously modeled as 
a black body \cite{1985MNRAS.217..105A},
a steep power law with $\alpha = 1-2$\cite{1993A&A...274..105W,bbf96},
relativistically smeared absorption above 0.6 keV \cite{2004MNRAS.349L...7G},
or as a collection of relativistically broadened
emission lines such as L shell transitions of Fe \cite{2006MNRAS.365.1067C}.
Tidal disruption events (TDEs) have X-ray spectra
that are very similar to those of NLS1s
but are transient, probably triggered when a star breaks up
after a passage close by a supermassive black
hole but are otherwise inactive \cite{2015JHEAp...7..148K}.

The thermal and emission line models would have 
coronal or disk-like geometries, with polarizations $\le$10\%.
There is growing evidence from rapid variability \cite{2020MNRAS.498..859D},
radio imaging \cite{2018A&A...614A..87B} that
NLS1s and some TDEs \cite{2018ApJ...856....1P,2020NewAR..8901538D}
have pc-scale relativistic jets in their cores, linking them to blazars.  Several NLS1s show evidence of collimated outflows contributing to the X-ray emission \cite{gallo18,2015MNRAS.449..129W}.
A detection of $>$10\% polarization would
favor a jet model of the soft excess, linking these sources to
blazars and rule out thermal and coronal models.
RE J1034$+$396 is an interesting case with a 3550 s
quasi-periodic oscillation \cite{2020arXiv200714704J}.

\subsection{Polarization metrics}

\label{sec:metrics}

The MDP is the lowest level of polarization that we can distinguish from
random noise at the 99\% confidence level. 
For an instrument with modulation factor $\mu$,
MDP $= 4.29/(\mu R T^{1/2}) (R+B)^{1/2}$\cite{weisskopf:77320E},
where $\mu$ is the ratio
of the polarization-modulated signal to the average,
$R$ is the source count rate, $T$ is the exposure time, and $B$ is the background rate.
When $B$ is small, as it is for our design (\S~\ref{sec:background}), then
MDP $= 4.29/(\mu [R T]^{1/2})$, so for a given exposure time and source flux density
MDP $\propto 1/(\mu {\mathcal A}^{1/2} )$, where ${\mathcal A} \equiv \int A d\lambda$ is sometimes referred to
as the integrated area of the system.
While MDP is a good indicator of instrument sensitivity, source modeling
requires good {\em measurements} of polarization, not just detections.
Thus, MDPs should be below the expected polarization fractions by $\times$2-3.

Sometimes PAs can be estimated from external data.  Then, the
fractional polarization can be estimated, as was
done effectively in POGO$+$ observations of Cyg X-1 to
constrain the lamppost model of the accretion disk corona \cite{2018NatAs...2..652C}.
In such a case, the uncertainty in the polarization fraction is
$\sigma_P = 1 / [\mu (0.5 R T )^{1/2}] \approx $ MDP$/3$ or 2-3\% for most of
our observations.

\subsection{Observing Plan}
\label{sec:obsplan}

\begin{figure}
   \centering
   \includegraphics[width=0.95\linewidth]{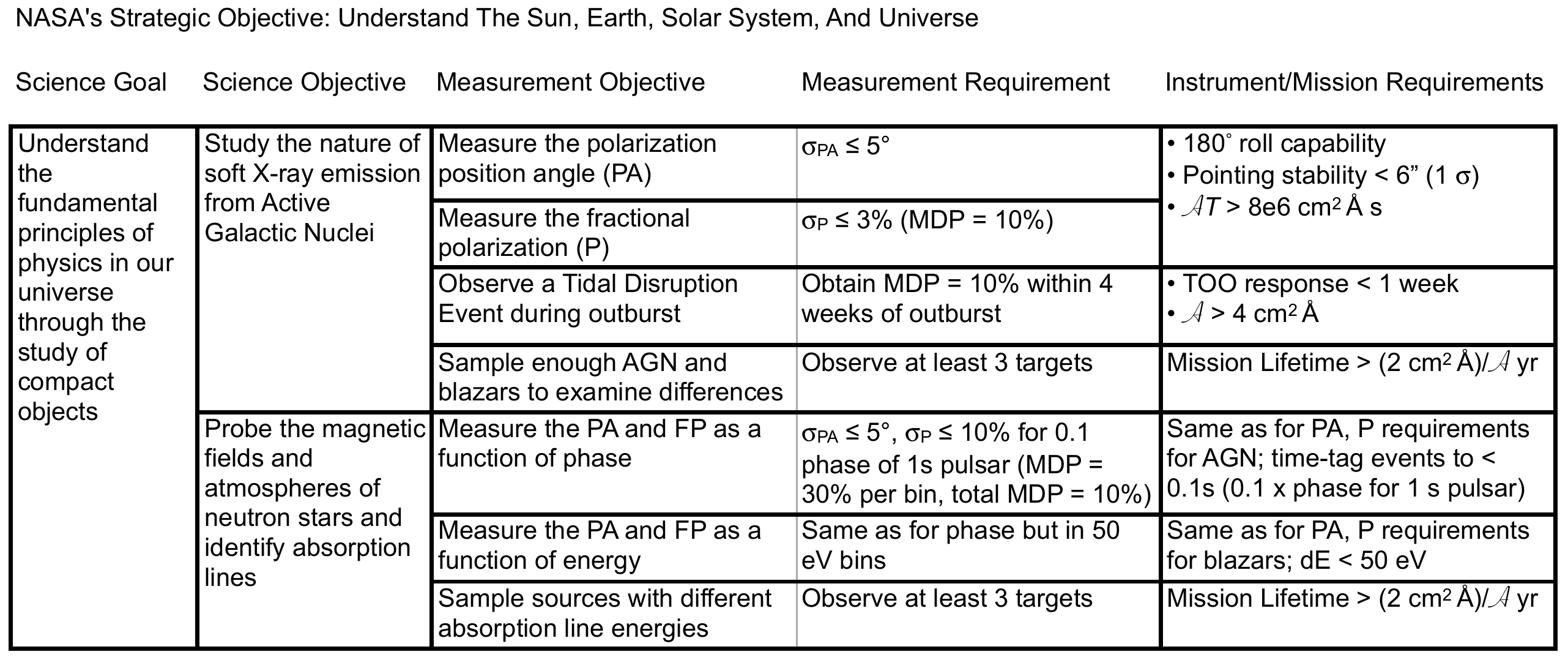}
\caption{\footnotesize
 \sxp\ science traceability matrix.
 $T$ is a target exposure time and $\mathcal{A}$ is the integrated area of the
 instrument (\S\ref{sec:metrics}).
 The minimum mission requires $\ge 3$ targets in both categories of source,
 requiring a mission lifetime $\ge$ (4 cm$^2$ \AA)$/\mathcal[a]$ yr; we have
 100\% margin with $\mathcal{A} = 8.3$ cm$^2$ \AA\ (\S\ref{sec:prediction}) and a 1 year mission.
 }
\label{fig:STM}
\end{figure}

A preliminary observing plan for a 1 year mission is given in table~\ref{tab:targets}
for an instrument with ${\mathcal A} = 8$ cm$^2$\AA, achievable with
our design (\S~\ref{sec:prediction}).
The plan includes 6 targets in each of our
primary categories: neutron stars and AGN with soft spectra.
Due to the difficulty of developing conclusions about an entire
class of source from one or two targets,
the ``minimum mission'' we consider scientifically useful is to
obtain $>$ 10\% MDP for each of 3 sources from both
of our two categories.  This criterion sets
the instrument requirements (Fig.~\ref{fig:STM}).
Published spectra for these sources were used to compute count rates in the \sxp\ instrument.
In addition to the science targets is a ``null'' calibrator -- a
bright, unresolved binary of cool stars with thermal coronae.
The table does not include the many other sources that
may be polarized and are comparably bright to the primary targets.  These
secondary targets include magnetic dwarf novae, nearby Galactic X-ray binaries
in outburst and other AGN that may be of
interest to guest observers after the first year of operation.

\begin{table}
\begin{center}
\caption{Preliminary Observing Plan$^{\rm a}$ \label{tab:targets} }
\begin{tabular}{lrrrr}
Category & Name &  Time & MDP  & Notes \\
	&   &	(Ms)	&	(\%)	 &	 \\
\hline
\hline
Solo NSs  & RX J1856.5$-$3754	&	1.0	&	8	&	Brightest isolated neutron star \\
    &   RX J0720.4$-$3125    &   1.0   &   10   &   Absorption line at 293 eV \cite{rxj0720,2012MNRAS.419.1525H} \\
    &   RX J1605.3$+$3249    &   1.5 &   9   &   Absorption line at 403 eV \cite{2012MNRAS.419.1525H} \\
    &   RX J1308.6$+$2127    &   1.0   &   10   &   Abs'n line varies: 107-256 eV \cite{2012MNRAS.419.1525H,2017MNRAS.468.2975B}  \\
XRB Pulsar  & Her X-1   &   0.3 &   3   &   Magnetosphere-disk interaction \\
Solo Pulsar & PSR B0656$+$14    & 1.0 &   7   &   Magnetospheric emission \\
\hline								
AGN &   Mk 421	&	0.1	&	3	&	Bright high spectral peak (HSP) blazar\\
    &   1H1426	&	0.5	&	5	&    HSP blazar\\
    &   Mk 501	&	0.5	&	5	&    HSP blazar\\
    &   3C 273	&	2.0 &   10	&	Low spectral peak blazar, soft excess\\
    &   RE J1034$+$396	&	1.0 &   8	&	Narrow Line Sy 1 (NLS1) \\
    &   Ark 564	&	1.0 &   8	&	NLS1 \\
    &   TBD	&	1.0 &   5	&	Tidal Disruption Event (TOO) \\
\hline								
Null & Capella	&	1.0	&	4	& Null polarization target \\
\hline
\end{tabular}
\end{center}
$^{\rm a}${\footnotesize
   We assume 40\% observing efficiency.  The blazars can be highly variable and may
   require substitution.}
\end{table}

\section{Payload Design}
\label{sec:payloaddesign}

The science instrument leverages previous funding various grants
to make focusing optics, multilayers, gratings, and CCDs.
Together, these high technical readiness level (TRL)
technologies enable us to build \sxp\ with minimal risk.
Schematically, the design is the same as for \redsox\ \cite{redsoxjatis}.
Here, we outline various modifications for \sxp.

 \begin{SCfigure}
    \centering
    \includegraphics[width=7.5cm]{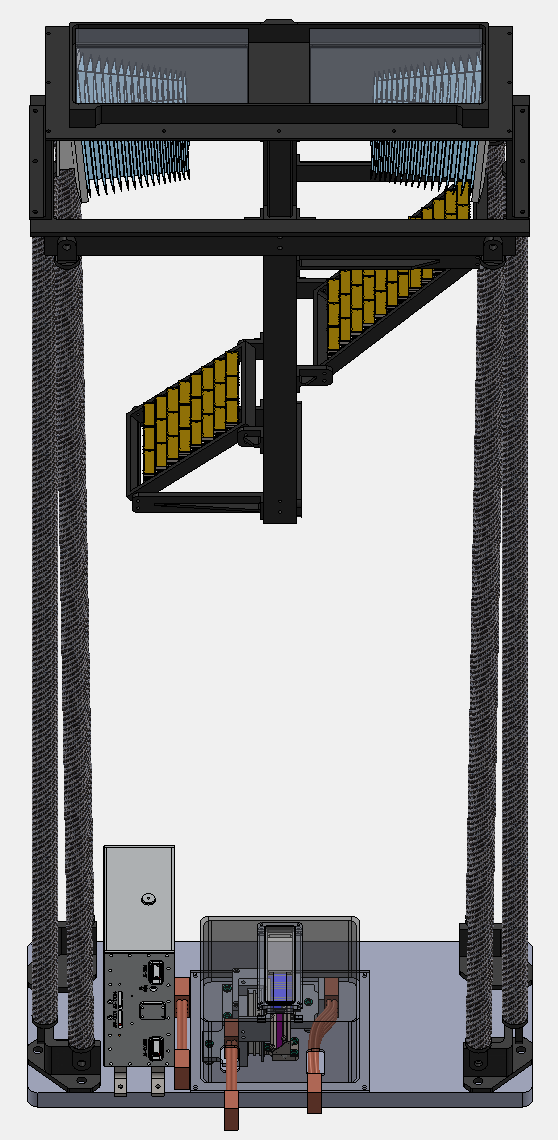}
 \caption{\footnotesize 
Cutaway rendering of the configuration for the \sxp\ instrument.
It is about 1.25~m tall, and has Si mirrors
made of curved and polished
segments about 100 mm tall, as developed by GSFC (colored light blue).
The optics section is 400 mm wide, 400 mm tall, and 100 mm deep, including the gratings (colored gold).
The orientation of the grating bars is parallel to the long dimension of the aperture
and gratings are blazed so that they disperse a focused spectrum at the laterally
graded multilayer mirror (LGML, magenta) inside the detector housing,
better shown in Fig.~\ref{fig:redsox_aperture}.
The detector housing is shown as a translucent
box but is designed to be light-tight with optical blocking
filters.  Only the imaging CCD detector (dark blue) is visible
in this rendering.  Thermal straps (copper) connect to the heat dissipation
system of the spacecraft.  Flexprints (not shown) connect the two detectors
in the detector housing to the tall box containing the digital and analog
electronics for driving the CCDs and finding events.
There is only one polarization
channel, so the instrument has to be rotated through 180\deg\ during
observations.  
More details of the optics section and the focal plane
are shown in Fig.~\ref{fig:redsox_aperture}. The payload $+z$ axis is along
the optical axis, with the origin at the imaging detector, and $+x$ is
toward and along the LGML.}
\label{fig:shortsxp}
\end{SCfigure}

\subsection{Focusing Optics}

\label{sec:optics}

The \sxp\ mirrors are
based on the Si metashell optics developed by the Goddard
Space Flight Center (GSFC) group under the direction
of Dr.\ William Zhang \cite{2019JATIS...5b1012Z}.
The GSFC group is currently using this technique to fabricate
a Wolter I mirror assembly for a sounding rocket payload, OGRE (Off-plane
Grating Rocket Experiment, \cite{ogel93}) led
by Prof.\ Randall McEntaffer (PSU).
The point spread function (PSF) of the mirror
assembly will have a half-power diameter (HPD) of $\lae 5$\arcsec.
The mirror assembly that will result from this work will
easily meet the science and spaceflight requirements of \sxp\ and will
have been tested before work on our optics would begin.

For \sxp, sources are only measurable
when on axis, so off-axis imaging quality is irrelevant.
Therefore, we baseline a purely parabolic, single-bounce
design for \sxp\ mirrors, eliminating hyperbolic secondaries.
A single bounce mirror
system requires half of the mirror segments and less
alignment work for a given effective area, making the
mirror system simpler, substantially less expensive, and much
lighter than a Wolter I design with the same geometric collecting area.
With Ni coating, reflectivities average 70\% for the graze
angles between 3 and 10\deg, giving a total effective
area of 130 cm$^2$ that is independent of energy
in the 0.2-0.4 keV band.

Because pure parabolic mirror systems are not common
in X-ray astronomy, we have studied the optical properties
of the images by raytracing.
We used the Interactive RayTrace (IRT) code for IDL
provided to MIT by Prof.\ Webster Cash (Colo. Univ.)
for the \axaf\ project to make spot diagrams, compute
HPDs, investigate off-axis aberrations, etc.
The telescope length, measured from the entrance
aperture to the imaging (and spectroscopy) focal plane
is limited to be no longer than
1.19 m due to the spatial limitations of the ESPA Grande
rideshare payload  (see \S~\ref{sec:scarchitecture}).
We will refer to this length as the
focal length of the system.

\begin{figure}
   \begin{center}
   \begin{tabular}{c}
   \includegraphics[width=9.cm]{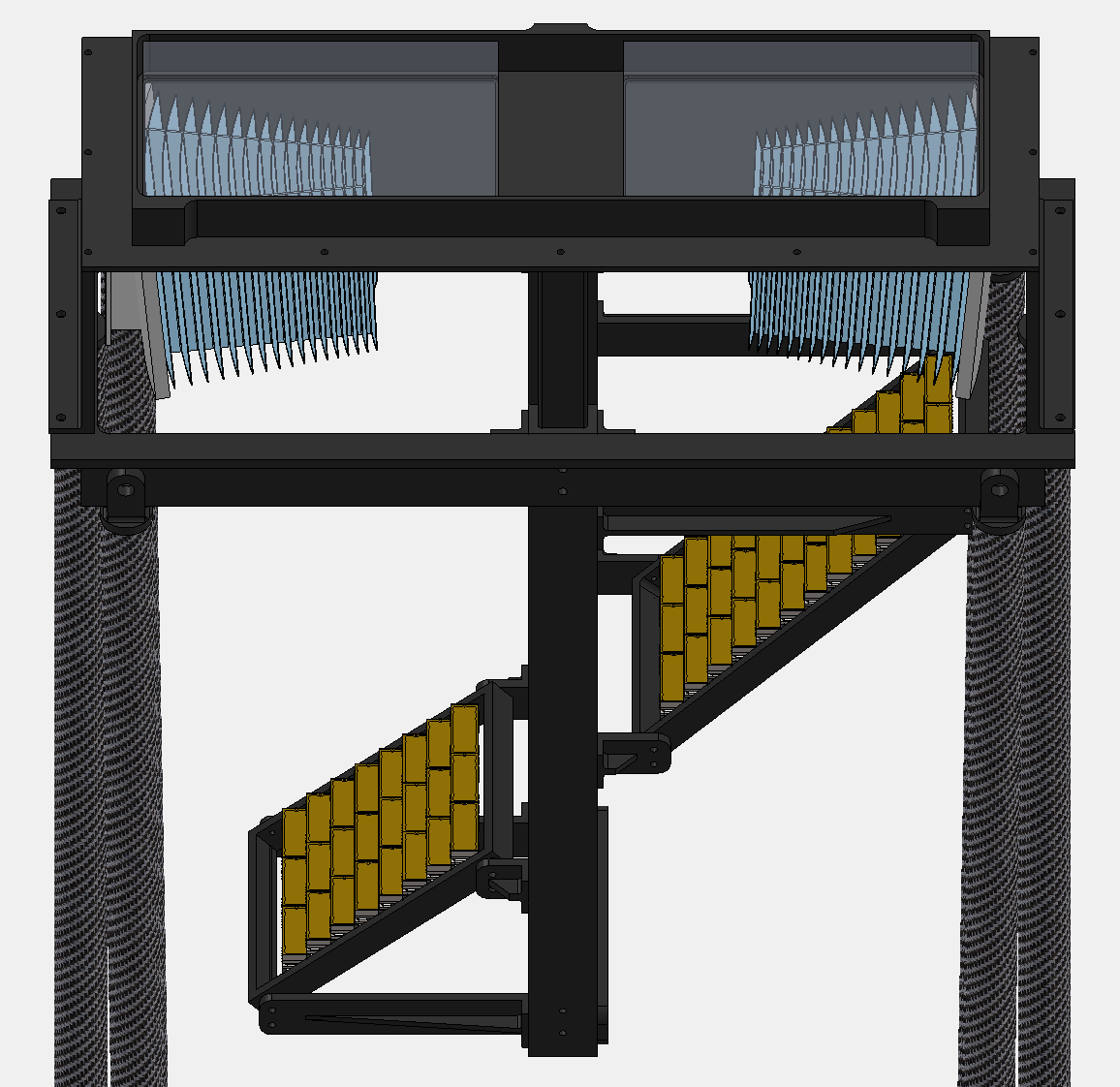}
    \includegraphics[width=6cm]{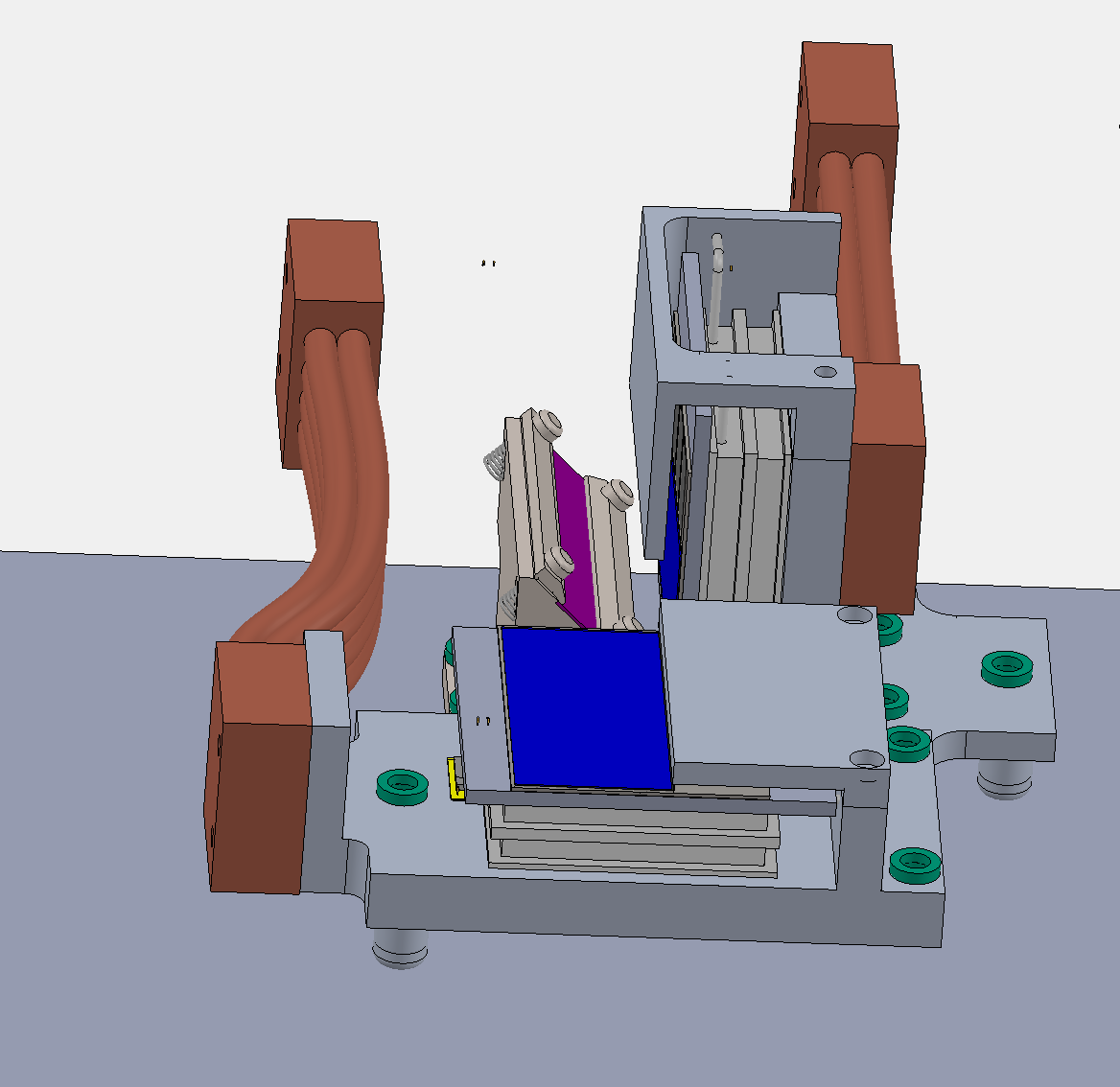}
 \end{tabular}
   \end{center}
   \vskip -0.2in
 \caption{\footnotesize
{\it Left:} View of \sxp\ optics section.  Thin Si mirrors (light blue) provide focusing.
    The gratings (gold) are stair-stepped to match the Bragg condition
	 at the multilayer mirror (see \S\ref{sec:catgratings} \cite{redsoxjatis}).
{\it Right:} Focal plane layout without the enclosing housing.
  The imaging CCD (dark blue) is at the direct focus while the LGML (magenta) is
  at the spectrometer focus and reflects to the vertically
  oriented polarimetry detector (dark blue).  Copper straps connect the
  CCD TECs to the spacecraft's cooling system.}
\label{fig:redsox_aperture}
   \vskip -0.2in
\end{figure}

The spectral resolution of a transmission grating spectrometer is
determined by the size of a point source's image along the dispersion.
Thus, only the 1D FWHM of the image is critical to the performance.
We find that azimuthally symmetric mirrors with a half-power diameter (HPD)
of $\lae$10\arcsec\ meet our requirements when sectored.
We use $\pm 18\deg$ sectors of mirror shells (Fig.~\ref{fig:redsox_aperture}),
on opposite sides of the optical axis
to reduce the telescope 2D PSF from 10\arcsec\ HPD to 2.9\arcsec\ in 1D (FWHM)
for use in spectroscopy \cite{1987ApOpt..26.2915C},
projecting to $\delta x = 17 \mu$m along the dispersion.
For a grating spectrometer, $R = \lambda/(\delta \lambda)
= x / (\delta x)$; $x = 11.4-35$ mm is the dispersion distance for the
gratings in our design (\S~\ref{sec:catgratings}) and the
baseline wavelength range of 30-65 \AA.
At these dispersion distances, $R$ is in the range 680 to 2000, a factor
of 7-20 better than the requirement.

\subsection{Gratings}

\label{sec:catgratings}

As with \redsox, we baseline
Critical Angle Transmission (CAT) gratings \cite{2009SPIE.7437E..14H}
developed in the Space Nanotechnology Lab (SNL) at MIT.
CAT gratings can be reliably
produced in a 10$\times$30 mm format \cite{2015SPIE.9603E..14H,heilmann17}.
Over a dozen CAT gratings of 200 nm period
and 4 $\mu$m membrane thickness have been tested in the
MIT polarimetry beamline,
including several in a 27 mm square format fabricated as part of
{\em Arcus} \cite{2017SPIE10397E..0QS} phase A development.
Efficiencies of 20\% are achieved in lab measurements
in first order as needed for \sxp\ (Garner, et al.\ in prep.).
Alignment is performed at the mount and assembly level,
as described in \S~\ref{sec:attitude}, based on a method
prototyped for {\em Arcus} \cite{2017SPIE10399E..15S}.
This method has been shown to align gratings to a single
reference to within the tolerances given in Table~\ref{tab:tolerance},
which are based on analysis and raytracing.

Precise positioning of the gratings
is given in our paper \cite{redsoxjatis} that includes raytrace validation.
For \sxp, there are 8 rows of 3 gratings
each in the lower sector and 12 rows of 3 each in the
upper sector for a total of 60 gratings.  Including 15\% blockage by
the gratings internal L1 and L2 support structure and 20\% blockage by
the grating assembly framework, the average transmission into $+1$ order
is 11.6\% across the \sxp\ band at the optimal blaze angle of 0.8\deg.

\subsection{Laterally Graded Multilayer Mirrors (LGMLs)}

\label{sec:lgmls}

LGMLs have been made by Eric Gullikson at the Lawrence
Berkeley National Lab (LBNL) Center for
X-ray Optics that are suitable for our design.
The Cr/B$_4$C/Sc layer combination has a reflectivity over 7\% between
31 and 70 \AA\ based on measurements at the Advanced
Light Source at LBNL \cite{2014SPIE.9144E..1KM,2015SPIE.9603E..19M}.

\subsection{CCD Detectors}

\label{sec:ccds}

The CCD detectors for PiSoX are designed and fabricated by the MIT Lincoln Laboratory (MITLL), which has provided CCD imaging detectors for X-ray missions dating back more than 25 years and including ASCA, \axaf, and Suzaku. The two CCID-94 devices to fly on PiSoX are backside-illuminated (BI), 50 mm $\times$25 mm (2048$\times$1024-pixel) frame-store CCDs that are equivalent to two side-by-side Suzaku CCID-41 devices, with 24-$\mu$m pixels. They trace direct heritage from the Suzaku BI CCD that performed exceptionally well for the nearly 10-year mission lifetime in a low-Earth, high-inclination orbit that regularly traversed the South Atlantic Anomaly, with features including a 3-phase polysilicon gate structure, low noise ($<$4 e$^-$) readout amplifiers, and charge injection.

One CCD will be used to record the dispersed, polarization-sensitive spectrum reflected by the LGML, and the second will record the zeroth-order source image to ensure proper acquisition of the target and monitor its broad-band flux,
giving us a redundant measurement of Stokes $I$.
Each CCD will be read out at 625 kHz through 8 parallel PJFET outputs, for a frame integration time of 1 s. The polarization detector will additionally have a ``parallel sum'' mode in which the imaging area is continuously read out and multiple rows are summed on-chip to provide time resolution of better than
100 ms
to allow phase-dependent analysis for
pulsars with periods of $>$ 1 s. For this reason, the dispersion direction is perpendicular to the CCD parallel transfer direction, and the dispersed spectrum will fit comfortably within the 50 mm width of the detector.

The CCD spectral resolution requirements are modest, with FWHM $<100$ eV required
to reduce background and to separate 0.2 keV photons from 0.4 keV photons from overlapping orders in the dispersed spectrum. This requirement plus the use of charge injection and short frame time allow operation at -50C with minimal dark current or CTI effects. This temperature will be controlled to $\pm$1C by one 2-stage thermo-electric cooler (TEC) on each CCD, removing heat to the spacecraft hot side via thermal straps. These high-heritage TECs are manufactured by II-VI Marlow, which fabricated similar TECs for Suzaku that worked flawlessly over the course of the 10-year mission.

Each CCD is coated with 50 nm of Al to block optical light, and the entire focal plane including the LGML is contained within a light-tight detector housing to further eliminate optical contamination from the bright Earth limb. The entrance aperture for the LGML and zeroth-order CCD is covered by a 50-nm Al + 50-nm polyimide filter mounted on a 95\% transparent metal mesh, manufactured by Luxel, which has provided similar filters for missions such as Chandra and Suzaku. This contamination blocking filter is kept warm (0--20C) with a heater to prevent molecular contamination from building up on the cold CCD surfaces or accumulating along other optical elements and reducing the soft X-ray response. The combined CCD QE and filter transmission is expected to be about 10\% at 0.2 keV and 50\% at 0.4 keV.

\subsection{Focal Plane Electronics}

\label{sec:electronics}

The focal plane electronics include three electronics boards that drive the CCDs and process the payload data for downlink.  The boards and their enclosure are based heavily on those flown on the Regolith X-Ray Imaging Spectrometer (REXIS), an instrument on board the OSIRIS-REx asteroid sample return mission.  The REXIS electronics performed nominally over a period of 3.5 years after launch. Two of the boards, the Interface
Board and Video Board, comprise the Detector Electronics (DE) and provide power and commanding to the detectors, provide precise clocking signals to the CCDs, and read out, digitize and process the detector signal. The existing REXIS design allows for processing of 16 parallel readout nodes, sufficient for the two 8-node CCDs aboard PiSoX, but will need to be modified to provide active thermal control.  The total power draw
is expected to be 20 W with 25\% margin.

The digitized raw frames are delivered to the third board, the Main Electronics Board (MEB), that includes a Xilinx Virtex 5 FPGA that further processes the raw image frames to greatly reduce the payload data volume. This digital processing follows the method used for over 20 years of X-ray missions, with on-board bias maps employed to identify and measure X-ray events, providing for each event the $x$ and $y$ position, a 3$\times$3 array of pulse heights, and a time counter.  Events associated with hot pixels or grades commonly associated with particle interactions can be rejected on-board. After processing, the MEB packages and transfers the event list and exposure information to the spacecraft electronics for downlink.  At 166 bits per event and a maximum likely count rate of 50 events per second between the two detectors, the maximum data rate is under 10 kbps. This includes 1 kbps devoted to instrument housekeeping. Occasionally, full CCD frames are needed to diagnose bias maps or an anomaly, requiring a downlink of 25 Mbit per CCD. In this diagnostic mode, consecutive frames need not be telemetered, and the full frame can be stored on-board and downlinked as possible.

\subsection{System Considerations and Other Components}

This section describes other aspects of the design of \sxp.
Table~\ref{tab:payload}
gives the payload component masses.

\begin{SCtable}
\caption{Payload Component Masses. 
 With 25\% margin, the total mass is 37.6 kg.  The focal
 plane includes two CCDs with associated TECs, the
 LGML and its mount, the optical blocking filters,
 thermal straps, and the detector housing.
 \label{tab:payload}}
\begin{tabular}{l|r}
Subsystem & Mass (kg)\\
\hline
\hline
Door Assembly &	5.5\\
Mirror Assembly &	7.4	\\
Grating Assembly &	3.4	\\
Focal Plane &	2.2	\\
Electrical Box &	4.3	\\
Optical Bench &	7.3\\
\hline
Total	&	30.1\\
\hline
\hline
\end{tabular}
\end{SCtable}

\subsubsection{Structure}

PiSoX will use a single-body rigid structure that is
as short as possible to fit into the ESPA
Grande payload volume (see \S~\ref{sec:scarchitecture}).
The payload consists of two major sections. The fore section, the optics module, contains
the Si shell mirror assembly and grating assembly.
The aft section (the focal plane) contains the detectors, multilayer mirror, and electronics systems.
Connecting these two sections are rigid struts made from carbon fiber reinforced polymer (CFRP).
Table~\ref{tab:tolerance} provides instrument subsystem tolerances
relative to the reference frame established by an optical cube on
the mirror assembly.
Tolerances were determined using analytical formulae
verified by raytracing \cite{redsoxjatis}.
Mechanical machining and assembly tolerances are sufficient
for most aspects of the system.
Aligning the gratings and the grating assemblies are carried out
on an optical bench and in the MIT X-ray Polarimetry beamline;
see sections \ref{sec:gratingintegration}
and \ref{sec:opticsintegration} for details.

\begin{table}
\begin{center}
\caption{Instrument subsystem tolerances (1 $\sigma$) \label{tab:tolerance} }
\begin{tabular}{l|ccc|ccc}
Subsystem & \multicolumn{3}{c}{Positional (mm)} & \multicolumn{3}{c}{Angular ($^\prime$)} \\
  & $x$ & $y$ & $z$ & $x$ & $y$ & $z$ \\
\hline
Grating assembly & \textcolor{OliveGreen}{1.0} & \textcolor{OliveGreen}{1.0} & \textcolor{OliveGreen}{1.0} &
 6.0 & 6.0 & \textcolor{OliveGreen}{60} \\
CAT grating in assembly & \textcolor{OliveGreen}{1.0} & \textcolor{OliveGreen}{1.0} & \textcolor{OliveGreen}{2.5} & \textcolor{OliveGreen}{60} & 6.0 & \textcolor{OliveGreen}{60} \\
ML mirror & 0.1 & \textcolor{OliveGreen}{1.0} & \textcolor{OliveGreen}{1.0} & 1 & \textcolor{OliveGreen}{60} & 6.0 \\
Imaging CCD & \textcolor{OliveGreen}{2.0} & \textcolor{OliveGreen}{5.0} & \textcolor{OliveGreen}{0.1} & \textcolor{OliveGreen}{120} & \textcolor{OliveGreen}{120} & \textcolor{OliveGreen}{120} \\
Polarimetry CCD & \textcolor{OliveGreen}{5.0} & \textcolor{OliveGreen}{2.0} & \textcolor{OliveGreen}{5.0} & \textcolor{OliveGreen}{120} & \textcolor{OliveGreen}{120} & \textcolor{OliveGreen}{120} \\
\hline
 \end{tabular}
\end{center}
{\small
   Tolerances are relative to a reference cube on the mirror assembly.
   Tolerances in green indicate those satisfied by machining and assembly accuracies
   to better than 0.1 mm; others will be achieved during alignment tests at MIT
   (\S\ref{sec:opticsintegration}).
   The coordinate system is defined in Fig.~\ref{fig:shortsxp}.
   }

\end{table}

\subsubsection{Attitude Control and Alignment}

\label{sec:attitude}

Based on optics ray tracing,
if pointing jitter is $\gae 20$\arcsec\ (3$\sigma$), light would be lost as the dispersed
spectrum is displaced from the Bragg peak of
the LGML along the dispersion, in a manner comparable to PSF broadening.
Therefore, we require that the spacecraft attitude determination control system (ADCS)
hold the orientation of the telescope's boresight to about 6\arcsec\ (1 $\sigma$).
In order to change the position angles of the LGML mirrors
with respect to the sky,
the spacecraft roll angle will be varied to span a range of 180\deg\ by continuously
rotating or stepping at
periodic intervals through the observation.
Due to possible source variability on time scales of 10 ks,
the satellite should complete a 180\deg\ scan at least once every 5000 s.
In order to accurately assign the roll angle to any given X-ray event, the scan rate
should be no faster than 1\deg/s and the roll angle and should
be accurate to 1\deg\ every second during data collection.
The initial roll angle for any given observation is arbitrary.
See \S~\ref{sec:scarchitecture} for more details about the
attitude system.

 \begin{SCfigure}
    \includegraphics[width=11cm]{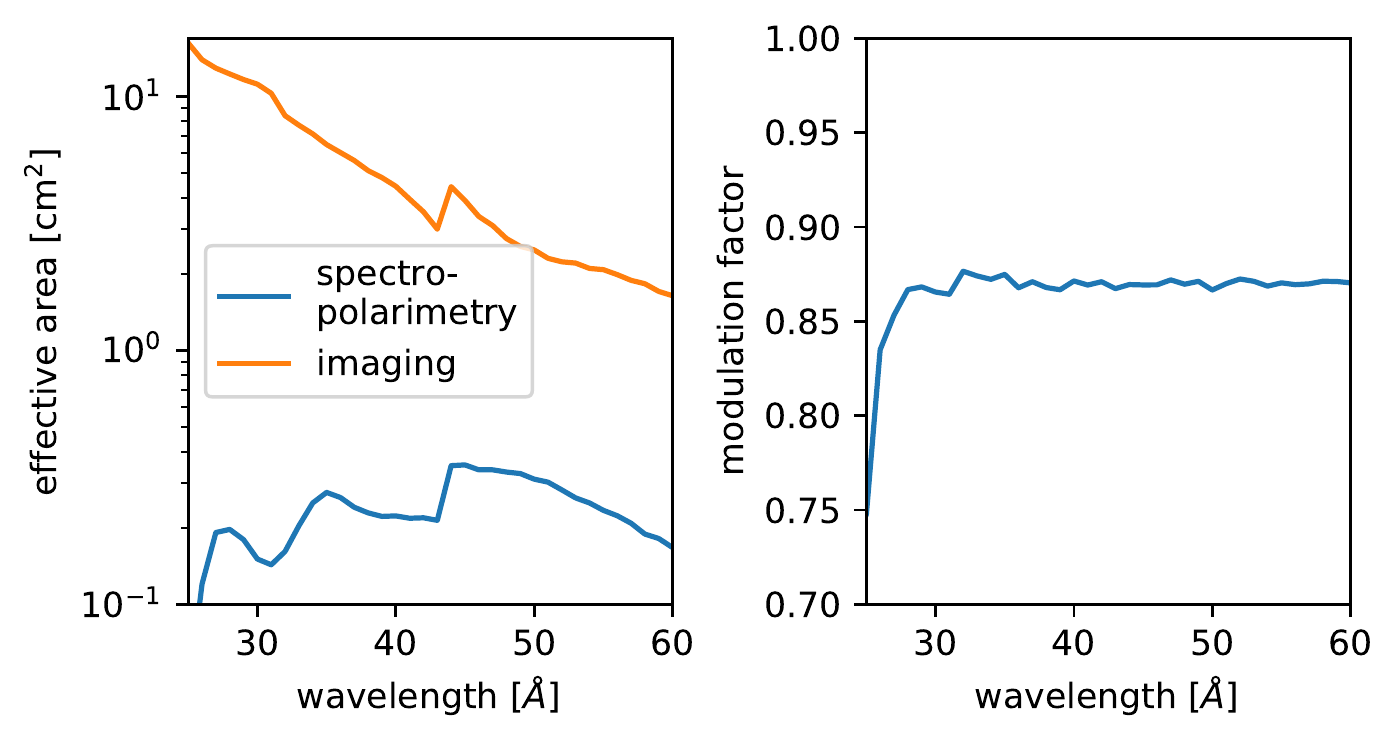}
 \caption{\footnotesize
    {\it Left:} Effective area of
  the system based on raytracing and confirmed by analysis.
  {\it Right:} Modulation factor, averaging
  about 87\%.
  }
\label{fig:mirrorhpd}

\end{SCfigure}

\subsubsection{Sources of Background}

\label{sec:background}

From the Suzaku mission (600 km altitude and 31\deg\ inclination),
the particle background in the
backside-illuminated CCD was
$5 \times 10^{-8}$ cnt/s/keV/pixel, or
$1.5 \times 10^{-5}$ cnt/s/mm$^2$ in a 0.2 keV band at 0.3 keV \cite{2008SPIE.7011E..2CL}.
In the cross dispersion direction, the dispersed spectra extraction region will be
about 3 mm tall due to the 20$\deg$ beam divergence
from the LGML, which is located at the spectroscopic focus.
In the dispersion direction, the spectra are 20 mm long giving a total of 60 mm$^2$ for
the extraction region.
For an exposure of $10^6$ s, we expect 900 counts from particles, while
obtaining an MDP of 10\% requires about 2,500 counts, so particle background
is minor but included in our MDP estimates in any case.
The X-ray background in the \sxp\ bandpass is dominated
by Galactic emission. A full ray-trace simulation including in particular gratings and the multi-layer mirror using the {\tt marxs} code \cite{2017AJ....154..243G,redsoxjatis} show that off-axis X-rays are dispersed in such a way that their reflection is suppressed by the ML, in contrast to on-axis photons which are diffracted such that they hit the Bragg-peak. Using a measured flux for the Galactic emission and background AGN\cite{2009PASJ...61.1117B}, the ray-trace predicts an X-ray background rate of 2 counts per 1 Ms -- clearly negligible. 

\subsubsection{Payload Technical Readiness Levels}

\label{sec:payloadtrl}

Si metashell optics are under fabrication at GSFC with APRA funding.
The entire process of cutting Si blocks, mounting into modules,
and then integrating into an assembly will have TRL $=$ 6 by Fall 2021.
Gratings have already been fabricated and tested that meet
our specifications, but assembly requires alignment that is currently
in the validation process using internal MIT funding.

As part of our current APRA grant to demonstrate technology for
soft X-ray polarimetry, MIT is acquiring a simple module of Si shell optics
from GSFC
to focus the X-rays from our electron impact
source within the MIT X-ray Polarimetry beamline
(see \S~\ref{sec:IandT}).
We expect to install the optics module in the beamline in the fall of 2021.
Thus,
we will be able to complete
an end-to-end test of the polarimeter design with all essential components on
rotational or translation stages as needed:
a 100\% polarized source, focusing optics, an internally aligned grating
assembly, a multilayer mirror, and the current lab CCD detector, raising the
TRL of the integrated system to 6.

\subsubsection{Baseline Performance}

\label{sec:prediction}

Using detailed analytical models of the reflectivity and QE curves and raytracing,
we obtain ${\mathcal A} = 8.3$ cm$^2$\AA,
consistent with the 1 year science plan (\S~\ref{sec:obsplan}).
The minimum mission (\S~\ref{sec:obsplan}) would still be possible if the
overall performance is degraded to ${\mathcal A} = 4$ cm$^2$\AA,
giving us about 108\% margin.
One may also estimate ${\mathcal A}$ by $A ~T_g \eta ~r ~Q ~\Delta \lambda$, where $A = 130$ cm$^2$
is the mirror effective area, $T_g = 0.116$ is the average grating efficiency including
obscuration by supports, $r=0.07$ is the
average reflectivity of the LGML, $Q=0.33$ is the average QE of the detector (including
the filter),
and $\Delta \lambda = 30$\AA\ is the bandwidth of the system.  The factor $\eta$ is
due to the combined effects of the optics broadening, attitude
variation, event centroiding, and structural variation during observations that
cause losses due to Bragg peak mismatch that we find to be
no more than 30\% ($\eta = 0.7$) based on raytracing.
This rough estimate gives ${\mathcal A} = 7.3$ cm$^2$\AA,
within 15\% of our detailed calculations.
The imaging detector
provides an independent, simultaneous measurement of $I$ in the 0.2-0.4 keV band
(and up to 10 keV),
with ${\mathcal A}_0 = 250$ cm$^2$\AA\ for zeroth order grating efficiencies
of about 20\%.
The zeroth order count rates for our targets are 0.2-20 cnt/s.

\subsection{Integration and Testing}

\label{sec:IandT}
\begin{figure}
    \centering
    \includegraphics[width=13cm]{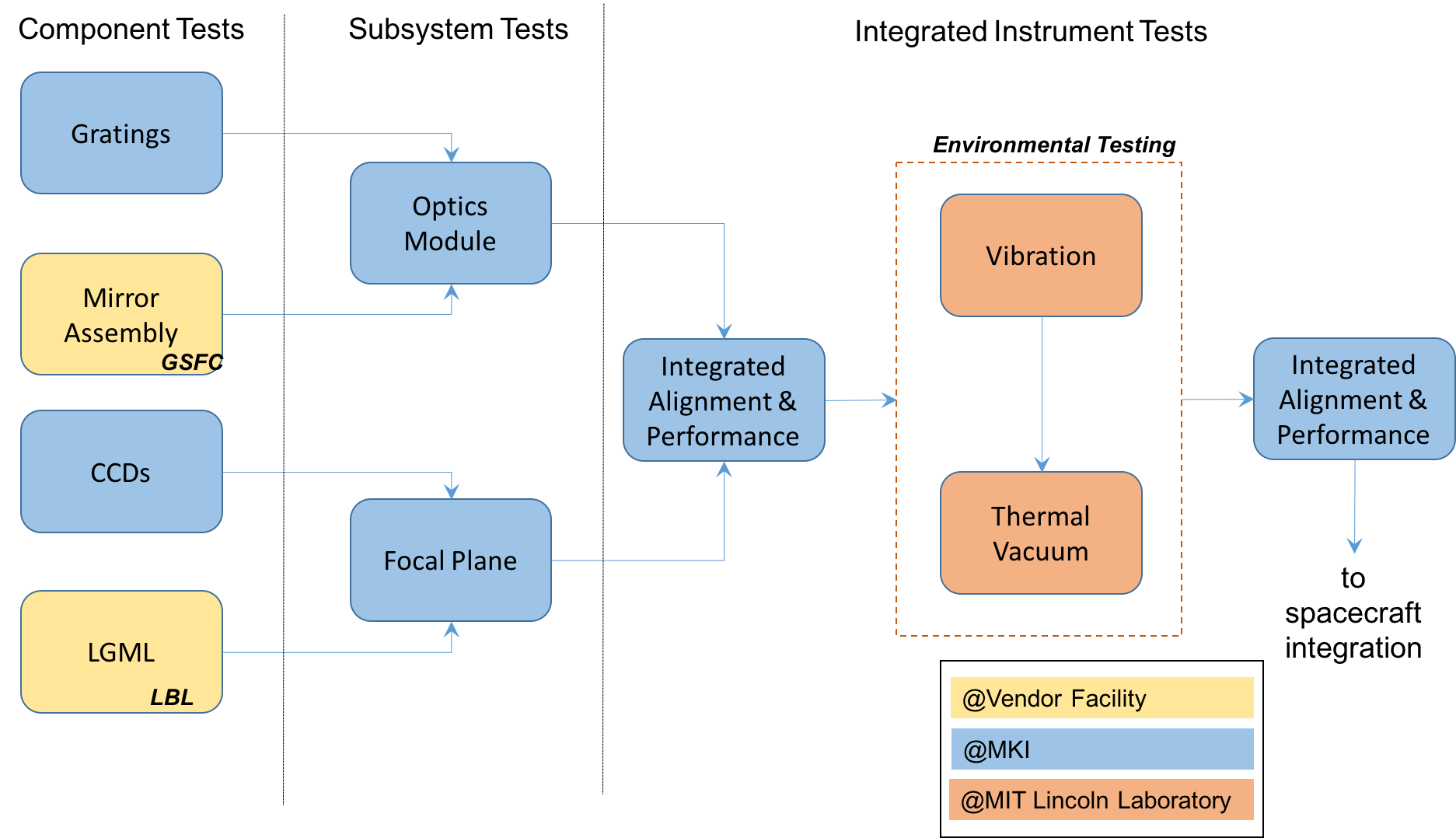}
 \caption{\footnotesize 
Block diagram of instrument-level integration and test.
GSFC and LBNL testing involve X-ray testing, as do all the MIT tests shown here.
The plan shows that all stages of development can be verified.  Both the MIT and LBNL
facilities use polarized X-ray sources.
}
\label{fig:aitPlan}

\end{figure}

Integration and test of the instrument will be the responsibility of the MIT payload team and will take place at MIT.  Subsystem testing, instrument integration, alignment and performance testing will occur at MIT using the MIT X-ray Polarimetry Laboratory beamline.  Environmental testing, including launch vibe and thermal vacuum testing, will take place at MITLL.  A high-level view of the integration and test plan is shown in Figure~\ref{fig:aitPlan}. The plan consists of three main sections: component testing, subsystem testing and integrated instrument tests.  Component-level testing of the mirror assembly and LGML will take place at GSFC and LBNL respectively, prior to delivery to MIT.  The gratings and the CCDs will be tested at MIT prior to subsystem integration.
At MIT, the gratings and the mirror assemblies are integrated into the optics module, and the CCDs and LGML are combined to form the focal plane.  Both of these subassemblies are tested separately, and then together at MIT in the Polarimetry beamline.  To complete instrument integration, the optics module and focal plane are assembled into the instrument structure.  Integrated alignment and performance testing is then conducted in the MIT beamline.  The instrument is then transported to MITLL for environmental testing, including vibration tests and thermal vacuum cycling and balance.  Upon the completion of environmental testing the instrument is returned to MIT for a final alignment and performance check in the X-ray beamline before delivery for integration with the spacecraft.

\subsubsection{MIT Facility}

\label{sec:beamline}

The MIT X-ray Polarimetry beamline is an 11m long X-ray beamline that is
extendable to 17m.  It was originally designed and constructed to test the
\axaf\ high energy transmission gratings.
It has since been adapted to produce polarized light at several energies in order to test components for soft X-ray polarimetry applications
and is described elsewhere
\cite{murphy:77322Y,2013SPIE.8861E..1DM,2014SPIE.9144E..1KM,2015SPIE.9603E..19M,garner19}.
Using the beamline, we will be able to separately test each component of \sxp, from optics through the focal plane.
Furthermore, we can test combinations of components such as the optics module, which is a combination
of the mirror and grating assemblies, and the focal plane, which has both flight CCDs and the LGML.
Finally, the assembled payload will fit into the 1.3m diameter chamber to complete
the performance and alignment testing.

\subsubsection{Component Testing: Gratings}
\label{sec:gratingintegration}
Testing of the performance and alignment of the CAT gratings is critical to \sxp\ and our team has the facilities and experience to do both. The performance and quality
of the gratings will be measured and verified in our X-ray beamline. These measurements have been performed before with similar CAT gratings, in particular using the prototype gratings for the Arcus mission (Garner, et al., these proceedings).

CAT grating alignment procedures have been developed and demonstrated
as part of Arcus phase A work \cite{2017SPIE10399E..15S,heilmann17}.
With an MIT internal grant, we have
built a grating alignment system modeled after this tool, built and tested for
CAT grating testing but customized for this project.
This technique uses a UV laser and position-sensitive detectors to measure the reflection and back diffraction off of the CAT grating bars.  From that information the orientation and alignment of the gratings can be determined within the tolerances required by \sxp\ (see \S\ref{tab:tolerance}).
An advantage of this method is that it can all be done on an optical bench. Given the number of gratings present in \sxp, this is a huge mitigation to schedule-related risk.
Using the laser-alignment tool, the \sxp\ CAT gratings will be aligned relative to one another as a complete subsystem and to an optical alignment cube mounted to the grating assembly.
The alignment will then be checked using the X-ray beamline (\S\ref{sec:beamline}).

\subsubsection{Subsystem Testing}

\label{sec:opticsintegration}
The \sxp\ optics module will be aligned and integrated in stages.
The grazing-incidence mirror assembly provided by GSFC will arrive as a unit aligned
to its own alignment cube.
The grating sub-assembly (described in \ref{sec:gratingintegration}) will then be aligned to the mirror assembly using the alignment cubes of the two subsystems on an
optical bench.
The MIT Polarimetry beamline can be used to verify alignment in X-rays for
both dispersed and 0th order imaging using the facility detector.
The shape and size of the images will provide a direct measurement of the alignment
when compared to images from raytraces that account for the finite source distance
and an incompletely illuminated aperture.
The finite source focus is about 130 mm shorter than the telescope
length, at 1.06 m from the telescope aperture.  A plate with
various 25 mm long slits can be scanned across the mirror entrance for
fine-grained testing.

The focal plane assembly, with both imaging and polarimetry detectors,
requires a determination of the LGML position as mounted on the focal plane with
respect to the imaging CCD.  ALS measurements provide accurate Bragg peak
positions with respect to an LGML edge that will be used as a reference
for mechanical assembly and metrology.  The positioning can be verified
using the MIT X-ray beamline as follows.  The LGML is scanned
in a monochromatic beam with a 0.1 mm slit to determine
where maximum reflectivity is achieved at 45\deg\ to the beamline.
By setting the X-ray target to different emission lines, the gradient
and the zero point relative to the LGML edge can be confirmed.  The stage is then
moved by the amount indicated for the zero point and X-ray beam will then
illuminate the detector reference point
that will provide a match to the Bragg peak for X-rays dispersed
by the gratings.

\subsubsection{Integrated Instrument Testing}

\label{sec:envirotesting}

After optics and focal plane subsystem integration, (\S\ref{sec:opticsintegration})
the two subsystems can be integrated with the optical bench and
tested in the MIT Polarimetry beamline without the facility detector.
The focal plane will be out of focus, but we can still use the 0th order image to
check for tip and tilt misalignment by comparing to predictions from raytraces.
The illumination of the LGML will be out of focus as well but the Bragg condition
will be met as long as the distance from the focal plane to the gratings is within
the required tolerance (see \S\ref{tab:tolerance}).  This will also allow us to test and
correct the relative positions of the focal plane and optics by adjusting the optical bench.
Critically, this approach allows us to do a full systems test of the payload that is accurate in every way except the distance to the source.

\section{Spacecraft}
\label{sec:spacecraft}
\subsection{Architecture}

\label{sec:scarchitecture}

\begin{SCfigure}
    \includegraphics[width=11cm]{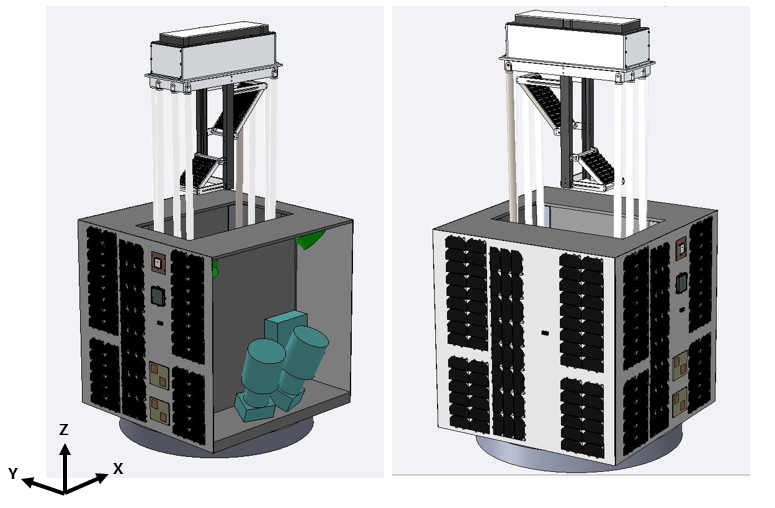}
 \caption{\footnotesize
  Renderings of the spacecraft with one external radiator panel around the star trackers hidden. Generally, the $+Z$ and $-Y$ faces are in shadow, while $+Y$ and $\pm X$ are used for power generation. 
   In both cases the sheathing and thermal layer
   around the payload are hidden.
  }
\label{fig:spacecraft}

\end{SCfigure}

The \sxp\ spacecraft is designed for low Earth orbit (LEO) and will be developed by NanoAvionics. 
Overviews of the structure and subsystems are provided in Figs.~\ref{fig:spacecraft}.
The platform will have subsystems from NanoAvionics and other companies
with high heritage in LEO, with a cold-redundant, dual string
architecture.
The structure will be composed of a 700 mm $\times$ 700 mm baseplate,
below which the Ruag PAS 610S ESPA ring separation system is mounted. 
The payload will be centrally mounted on top of this base plate
and shear panels will be mounted around the instrument
with most of the electronics installed on them. 

\subsection{Electronics \& Power}

The flight computer (FC) is the NanoAvionics SatBus 3C2, running the FreeRTOS real time
operating system on an ARM 32-bit Cortex M7 CPU with a clock speed of 400 MHz,
double-precision floating point unit, 1 MB of internal RAM, 2MB of internal FLASH
memory, as well as 3 internal magnetometers.
An identical FC is hot-swappable in flight.
The FC interfaces to the 
NanoAvionics Payload Controller (or its backup), which runs Linux, has
256 MB of data storage via microSD cards, and communicates with
the payload electronics via TCP/IP.

\subsection{Attitude Control}

Attitude knowledge and control is handled by the
attitude determination and control system (ADCS) on the spacecraft
bus.
Components are commercially
available with substantial flight heritage.
Attitude knowledge and control requirements are
20\arcsec{} at 3 $\sigma$ in order to maintain
the alignment of the dispersed spectrum and the Bragg
peak of the LGML (see \S\ref{sec:attitude}).
There are two aspects of the attitude control budget: short term
variations over seconds to hours and fixed offsets.
The former are not correctable
in flight so they contribute to the pointing error.  The
latter will be calibrated out in flight during the
commissioning phase of the mission that begins when the
telescope door is opened.

\subsection{Communications}

Three communications channels are available, all with redundant
components.
A link budget was established for each of UHF, S-Band and X-Band
assuming an orbital altitude of 550 km.
X-band can be used in ground station tracking mode with a 5m dish
for all elevations above 5\deg.

\section{Mission Implementation}
\label{sec:mission}

\subsection{Launch Vehicle and Orbit}

\label{sec:launchandorbit}

\begin{SCfigure}
    \centering
    \includegraphics[width=9cm]{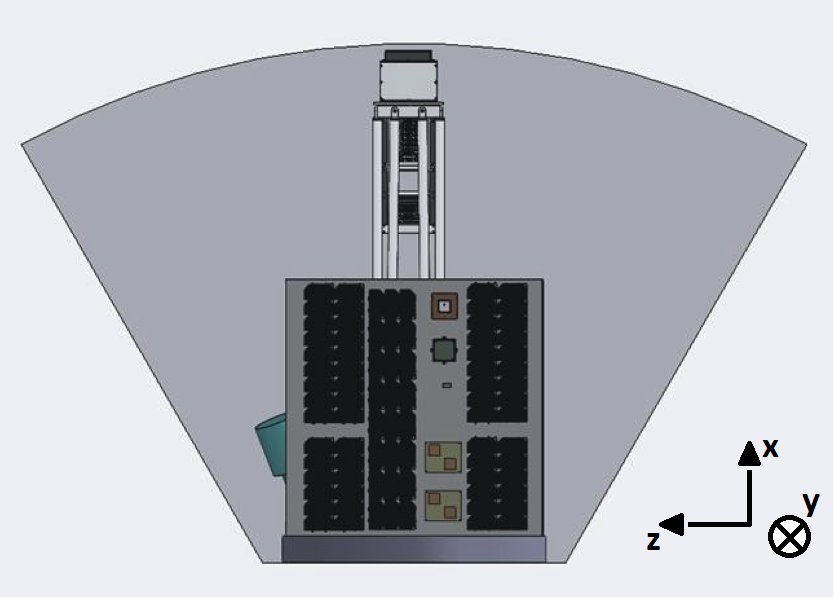}
 \caption{\footnotesize 
Rendering of the spacecraft in the Starlink rideshare volume.
The dimensions of the satellite are $720 \times 735 \times 1400$ mm,
within the payload accommodation
volume (\S\ref{sec:launchandorbit}).
\sxp\ also conforms to the ESPA Grande payload
requirements on mass and center of gravity (\S\ref{sec:launchandorbit}).
}
\label{fig:spacecraftespa}
\end{SCfigure}

The mission is designed for operations in LEO for a
lifetime of one year.
An injection altitude of $>$ 350 km will provide this orbital lifetime
for \sxp, which has a mass to area ratio of about $210$ km/m$^2$,
While low inclinations (equatorial orbits) are preferred
due to reduced particle
background, moderate inclinations are acceptable,
such as when launching from Florida.
The satellite meets ESPA Grande Payload Rideshare requirements,
(42\arcsec{} $\times$ 46\arcsec{} $\times$ 56\arcsec{}, or
$1067 \times 1168 \times 1422$ mm),
as can be seen in Figure \ref{fig:spacecraftespa}.

The satellite mass estimate is 148.0 kg, including a 10\%
mass margin on top of 5-10\% subsystem mass margin
and the 25\% mass margin on the instrument.
The ESPA interface places an upper mass limit of 465kg on the spacecraft.
Thus, we have over 200\% mass margin for the ESPA
interface. 
The satellite center of gravity is 16.7\arcsec{} from
the ESPA ring interface and 0.32\arcsec{} from the ring
center-line, satisfying the requirements
for the ESPA Grande interface to be less than 20\arcsec{} from
the interface and less than 1\arcsec{} from the center-line in
both other directions.

\subsection{Operations}
\label{sec:ops}

Given the long exposures, data download is only required once per week via
X-band.
At the maximum instrument data rate and 40\% observing efficiency,
a week of data amounts to
about 300 MB, easily downloaded in a few minutes via X-band.
Operations are based on one S-band uplink per day
and one X-band contact per week.
Alternatively, data can be downloaded via S-band in 2-3 passes per day.
The NanoAvionics Operations Center (NOC) in Vilnius
is responsible for scheduling contacts, uploading
commands, and storing downlinked data onto a secure server.
Data will then be available
for download to the \sxp\ Science Operations
Center (PSOC) at MIT.
The PSOC will be responsible for long-term observing
plan that is used to create the detailed
observing plan on a monthly basis.
The PSOC will collect data from the NOC and convert housekeeping and
engineering data to physical units for assessing the health of the
instrument.

\section{Technology Development}

Several systems of \sxp\ instrument would be flown in orbit for the first time on this mission
and are vital to future missions.  Thus, \sxp\ would raise the TRLs of these components (see \S\ref{sec:payloadtrl}).

Si metashell mirrors are lightweight and have the potential to provide
sub-arcsecond imaging. 
While Si metashell mirrors are being developed for OGRE, \sxp\ would provide the first
long-term use of these optics in Earth orbit.

CAT gratings are currently planned for Arcus \cite{2017SPIE10397E..0QS,heilmann17},
proposed as a MidEx mission and as an option for Lynx \cite{heilmann17}.
Arcus will provide high-resolution soft X-ray spectroscopy in the 12-50\AA\ bandpass with sensitivity orders of magnitude higher than any previous astronomical observatory. 
Lynx is a large X-ray imaging telescope being
considered as a flagship mission that would follow on from the successes of \axaf\ \cite{lynx,2019SPIE11118E..0JB}.

The CCID-94 detectors (\S~\ref{sec:ccds}) are a straightforward evolution of
the CCID-41 devices flown successfully on Suzaku and have
design heritage in \axaf, Hete-2 and ASCA.
Front-illuminated versions of these devices were successfully fabricated during Arcus \cite{2017SPIE10397E..0QS} Phase A.
CCID-94 chips  are currently planned for Arcus and Lynx.
The \sxp\ project builds on the development effort
for these missions that has been underway
at MIT and MITLL for many years.

Laterally graded multilayer mirrors (LGMLs, \S~\ref{sec:lgmls}) have never been flown and are the critical
polarizing element of our design, which is the only broad-band polarimetry
method in the sub-keV band of which we are aware.
Just as \ixpe\ is an Explorer to measure polarization in the 2-8 keV band,
\sxp\ is a Pioneer of the sub-keV band.
The X-ray Polarization Probe (XPP, \cite{xpp}) would
combine larger scale versions of these instruments
with a 10-50 keV polarimeter, such as demonstrated by
X-Calibur \cite{10.1117/1.JATIS.4.1.011004}.
Thus, XPP would be a general observatory, capable of measuring polarizations
across the 0.1-50 keV band simultaneously.
Currently funded lab development is focused on extending LGML usage up to 0.8 keV,
more thoroughly covering the sub-keV band.
See the XPP white paper \cite{xpp} and the accompanying
science white paper \cite{2019arXiv190409313K} for more details.

\acknowledgments 

Support for this work was provided in part through NASA grants
80NSSC20K1249, NNH18ZDA001N, and NNX15AL14G.
Laboratory supplies and equipment were funded in part through the
MIT Kavli Research Investment Fund.
Support for this work was also provided in part through NASA grant NNX17AG43G and
Smithsonian Astrophysical Observatory (SAO) contract SV3-73016 to MIT
for support of the {\em Chandra} X-Ray Center (CXC), which is operated
by SAO for and on behalf of NASA under contract NAS8-03060.

\bibliographystyle{spiebib} 

\begin{thebibliography}{10}

\bibitem{ixpe}
{O'Dell}, S.~L., {Baldini}, L., {Bellazzini}, R., {Costa}, E., {Elsner}, R.~F.,
  {Kaspi}, V.~M., {Kolodziejczak}, J.~J., {Latronico}, L., {Marshall}, H.~L.,
  {Matt}, G., {Mulieri}, F., {Ramsey}, B.~D., {Romani}, R.~W., {Soffitta}, P.,
  {Tennant}, A.~F., {Weisskopf}, M.~C., {Allen}, D.~Z., {Amici}, F.,
  {Antoniak}, S., {Attina}, P., {Bachetti}, M., {Barbanera}, M., {Baumgartner},
  W.~H., {Bladt}, J., {Bongiorno}, S.~D., {Borotto}, F., {Brooks}, T.,
  {Bussinger}, S., {Bygott}, H.~K., {Cavazzuti}, E., {Ceccanti}, M., {Citraro},
  S., {Deininger}, W.~D., {Del Monte}, E., {Dietz}, K.~L., {Di Lalla}, N., {Di
  Persio}, G., {Donnarumma}, I., {Erickson}, J., {Evangelista}, Y., {Fabiani},
  S., {Ferrazzoli}, R., {Foster}, M., {Giusti}, M., {Gunji}, S., {Guy}, L.,
  {Johnson}, S., {Kalinowski}, W., {Kelley}, A.~R., {Kilaru}, K., {Lefevre},
  C., {Maldera}, S., {Manfreda}, A., {Marengo}, M., {Masciarelli}, J.,
  {McEachen}, M., {Mereu}, P., {Minuti}, M., {Mitchell}, M.~A., {Mitchell}, S.,
  {Mitsuishi}, I., {Morbidini}, A., {Mosti}, F., {Nasimi}, H., {Negri}, B.,
  {Orsini}, L., {Osborne}, D., {Pavelitz}, S.~D., {Pentz}, C., {Perri}, M.,
  {Pesce-Rollins}, M., {Peterson}, C., {Piazzolla}, R., {Pieraccini}, S.,
  {Pilia}, M., {Pinchera}, M., {Puccetti}, S., {Ranganathan}, J., {Read}, T.,
  {Rubini}, A., {Santoli}, F., {Sarra}, P., {Schindhelm}, S., {Sciortino}, A.,
  {Seckar}, C., {Sgr{\`o}}, C., {Smith}, B.~T., {Speegle}, C.~O., {Tamagawa},
  T., {Tardiola}, M., {Tobia}, A., {Tortosa}, A., {Trois}, A., {Weddendorf},
  B., {Wedmore}, J., and {Zanetti}, D., ``{The Imaging X-ray Polarimetry
  Explorer (IXPE): technical overview},'' in [{\em
  \procspie}{\nolinebreak\hspace{0.1em}]},  {\em Society of Photo-Optical
  Instrumentation Engineers (SPIE) Conference Series} {\bf 10699},  106991X
  (Aug 2018).

\bibitem{1985ApJ...297..621A}
{Antonucci}, R.~R.~J. and {Miller}, J.~S., ``{Spectropolarimetry and the nature
  of NGC 1068},'' {\em \apj}~{\bf 297},  621--632 (Oct. 1985).

\bibitem{1959AJ.....64Q.339M}
{Mayer}, C.~H. and {Sloanaker}, R.~M., ``{Polarization of the 10-cm radiation
  from the Crab Nebula and other sources.},'' {\em \aj}~{\bf 64},  339 (Jan.
  1959).

\bibitem{2019ARA&A..57..467B}
{Blandford}, R., {Meier}, D., and {Readhead}, A., ``{Relativistic Jets from
  Active Galactic Nuclei},'' {\em \araa}~{\bf 57},  467--509 (Aug 2019).

\bibitem{radha}
{Radhakrishnan}, V., {Cooke}, D.~J., {Komesaroff}, M.~M., and {Morris}, D.,
  ``{Evidence in Support of a Rotational Model for the Pulsar PSR 0833-45},''
  {\em \nat}~{\bf 221},  443--+ (1969).

\bibitem{2006RPPh...69.2631H}
{Harding}, A.~K. and {Lai}, D., ``{Physics of strongly magnetized neutron
  stars},'' {\em Reports on Progress in Physics}~{\bf 69},  2631--2708 (Sep
  2006).

\bibitem{2008AIPC..983..340H}
{Ho}, W.~C.~G. and {Mori}, K., ``{Modeling Phase-resolved Observations of the
  Surfaces of Magnetic Neutron Stars},'' in [{\em 40 Years of Pulsars:
  Millisecond Pulsars, Magnetars and More}{\nolinebreak\hspace{0.1em}]},
  {Bassa}, C., {Wang}, Z., {Cumming}, A., and {Kaspi}, V.~M., eds., {\em
  American Institute of Physics Conference Series} {\bf 983},  340--344 (Feb.
  2008).

\bibitem{2010A&A...522A.111S}
{Suleimanov}, V., {Hambaryan}, V., {Potekhin}, A.~Y., {van Adelsberg}, M.,
  {Neuh{\"a}user}, R., and {Werner}, K., ``{Radiative properties of highly
  magnetized isolated neutron star surfaces and approximate treatment of
  absorption features in their spectra},'' {\em \aap}~{\bf 522},  A111 (Nov.
  2010).

\bibitem{2002Natur.420...51C}
{Cottam}, J., {Paerels}, F., and {Mendez}, M., ``{Gravitationally redshifted
  absorption lines in the X-ray burst spectra of a neutron star},'' {\em
  \nat}~{\bf 420},  51--54 (Nov. 2002).

\bibitem{2008ApJ...672..504C}
{Cottam}, J., {Paerels}, F., {M{\'e}ndez}, M., {Boirin}, L., {Lewin}, W.~H.~G.,
  {Kuulkers}, E., and {Miller}, J.~M., ``{The Burst Spectra of EXO 0748-676
  during a Long 2003 XMM-Newton Observation},'' {\em \apj}~{\bf 672},  504--509
  (Jan. 2008).

\bibitem{rxj0720}
{Haberl}, F., {Zavlin}, V.~E., {Tr{\"u}mper}, J., and {Burwitz}, V., ``{A
  phase-dependent absorption line in the spectrum of the X-ray pulsar RX
  J0720.4-3125},'' {\em \aap}~{\bf 419},  1077--1085 (June 2004).

\bibitem{2007ApSS.308..181H}
{Haberl}, F., ``{The magnificent seven: magnetic fields and surface temperature
  distributions},'' {\em \apss}~{\bf 308},  181--190 (Apr. 2007).

\bibitem{2017MNRAS.468.2975B}
{Borghese}, A., {Rea}, N., {Coti Zelati}, F., {Tiengo}, A., {Turolla}, R., and
  {Zane}, S., ``{Narrow phase-dependent features in X-ray dim isolated neutron
  stars: a new detection and upper limits},'' {\em \mnras}~{\bf 468},
  2975--2983 (July 2017).

\bibitem{2012MNRAS.419.1525H}
{Hohle}, M.~M., {Haberl}, F., {Vink}, J., {de Vries}, C.~P., and
  {Neuh{\"a}user}, R., ``{Narrow absorption features in the co-added XMM-Newton
  RGS spectra of isolated neutron stars},'' {\em \mnras}~{\bf 419},  1525--1536
  (Jan. 2012).

\bibitem{2009A&A...500..891S}
{Suleimanov}, V., {Potekhin}, A.~Y., and {Werner}, K., ``{Models of magnetized
  neutron star atmospheres: thin atmospheres and partially ionized hydrogen
  atmospheres with vacuum polarization},'' {\em \aap}~{\bf 500},  891--899 (Jun
  2009).

\bibitem{2012A&A...546A.121P}
{Potekhin}, A.~Y., {Suleimanov}, V.~F., {van Adelsberg}, M., and {Werner}, K.,
  ``{Radiative properties of magnetic neutron stars with metallic surfaces and
  thin atmospheres},'' {\em \aap}~{\bf 546},  A121 (Oct. 2012).

\bibitem{2002PhRvD..66b3002H}
{Heyl}, J.~S. and {Shaviv}, N.~J., ``{QED and the high polarization of the
  thermal radiation from neutron stars},'' {\em \prd}~{\bf 66},  023002--+
  (July 2002).

\bibitem{hsl03}
Heyl, J.~S., Shaviv, N.~J., and Lloyd, D., ``The high-energy
  polarization-limiting radius of neutron star magnetospheres - i. slowly
  rotating neutron stars,'' {\em Monthly Notices of the Royal Astronomical
  Society}~{\bf 342}(1),  134--144 (2003).

\bibitem{2015MNRAS.454.3254T}
{Taverna}, R., {Turolla}, R., {Gonzalez Caniulef}, D., {Zane}, S., {Muleri},
  F., and {Soffitta}, P., ``{Polarization of neutron star surface emission: a
  systematic analysis},'' {\em \mnras}~{\bf 454},  3254--3266 (Dec 2015).

\bibitem{2019A&A...622A..61S}
{Staubert}, R., {Tr{\"u}mper}, J., {Kendziorra}, E., {Klochkov}, D., {Postnov},
  K., {Kretschmar}, P., {Pottschmidt}, K., {Haberl}, F., {Rothschild}, R.~E.,
  {Santangelo}, A., {Wilms}, J., {Kreykenbohm}, I., and {F{\"u}rst}, F.,
  ``{Cyclotron lines in highly magnetized neutron stars},'' {\em \aap}~{\bf
  622},  A61 (Feb 2019).

\bibitem{meszaros88}
{Meszaros}, P., {Novick}, R., {Szentgyorgyi}, A., {Chanan}, G.~A., and
  {Weisskopf}, M.~C., ``{Astrophysical implications and observational prospects
  of X-ray polarimetry},'' {\em \apj}~{\bf 324},  1056--1067 (Jan. 1988).

\bibitem{2005ApJ...633.1064H}
{Hickox}, R.~C. and {Vrtilek}, S.~D., ``{Pulse-Phase Spectroscopy of SMC X-1
  with Chandra and XMM-Newton: Reprocessing by a Precessing Disk?},'' {\em
  \apj}~{\bf 633},  1064--1075 (Nov. 2005).

\bibitem{chandra60}
{Chandrasekhar}, S.,  [{\em {Radiative transfer}}{\nolinebreak\hspace{0.1em}]},
  Dover Publications Inc: New York (1960).

\bibitem{1980ApJ...235..386M}
{Marscher}, A.~P., ``{Relativistic jets and the continuum emission in QSOs},''
  {\em \apj}~{\bf 235},  386--391 (Jan. 1980).

\bibitem{2004ApJ...600..127G}
{Giroletti}, M., {Giovannini}, G., {Feretti}, L., {Cotton}, W.~D., {Edwards},
  P.~G., {Lara}, L., {Marscher}, A.~P., {Mattox}, J.~R., {Piner}, B.~G., and
  {Venturi}, T., ``{Parsec-Scale Properties of Markarian 501},'' {\em
  \apj}~{\bf 600},  127--140 (Jan. 2004).

\bibitem{1985ApJ...298..114M}
{Marscher}, A.~P. and {Gear}, W.~K., ``{Models for high-frequency radio
  outbursts in extragalactic sources, with application to the early 1983
  millimeter-to-infrared flare of 3C 273},'' {\em \apj}~{\bf 298},  114--127
  (Nov. 1985).

\bibitem{1999AJ....117.2185P}
{Perlman}, E.~S., {Biretta}, J.~A., {Zhou}, F., {Sparks}, W.~B., and
  {Macchetto}, F.~D., ``{Optical and Radio Polarimetry of the M87 Jet at 0.2''
  Resolution},'' {\em \aj}~{\bf 117},  2185--2198 (May 1999).

\bibitem{2016ApJ...832....3A}
{Avachat}, S.~S., {Perlman}, E.~S., {Adams}, S.~C., {Cara}, M., {Owen}, F.,
  {Sparks}, W.~B., and {Georganopoulos}, M., ``{Multi-wavelength Polarimetry
  and Spectral Study of the M87 Jet During 2002-2008},'' {\em \apj}~{\bf 832},
  3 (Nov 2016).

\bibitem{2002ApJ...564..683M}
{Marshall}, H.~L., {Miller}, B.~P., {Davis}, D.~S., {Perlman}, E.~S., {Wise},
  M., {Canizares}, C.~R., and {Harris}, D.~E., ``{A High-Resolution X-Ray Image
  of the Jet in M87},'' {\em \apj}~{\bf 564},  683--687 (Jan 2002).

\bibitem{gallo18}
{Gallo}, L., ``{X-ray perspective of Narrow-line Seyfert 1 galaxies},'' in
  [{\em Revisiting Narrow-Line Seyfert 1 Galaxies and their Place in the
  Universe}{\nolinebreak\hspace{0.1em}]},   34 (Apr. 2018).

\bibitem{1985MNRAS.217..105A}
{Arnaud}, K.~A., {Branduardi-Raymont}, G., {Culhane}, J.~L., {Fabian}, A.~C.,
  {Hazard}, C., {McGlynn}, T.~A., {Shafer}, R.~A., {Tennant}, A.~F., and
  {Ward}, M.~J., ``{EXOSAT observations of a strong soft X-ray excess in MKN
  841.},'' {\em \mnras}~{\bf 217},  105--113 (Nov. 1985).

\bibitem{1993MNRAS.261..346H}
{Haardt}, F. and {Matt}, G., ``{X-ray polarization in the two-phase model for
  AGN and X-ray binaries},'' {\em \mnras}~{\bf 261},  346--352 (Mar. 1993).

\bibitem{2012MNRAS.426L.101M}
{Marin}, F., {Goosmann}, R.~W., {Dov{\v c}iak}, M., {Muleri}, F., {Porquet},
  D., {Grosso}, N., {Karas}, V., and {Matt}, G., ``{X-ray polarimetry as a new
  tool to discriminate reflection from absorption scenarios - predictions for
  MCG-6-30-15},'' {\em \mnras}~{\bf 426},  L101--L105 (Oct. 2012).

\bibitem{1993A&A...274..105W}
{Walter}, R. and {Fink}, H.~H., ``{The ultraviolet to soft X-ray bump of
  Seyfert 1 type active galactic nuclei.},'' {\em \aap}~{\bf 274},  105 (July
  1993).

\bibitem{bbf96}
{Boller}, T., {Brandt}, W.~N., and {Fink}, H., ``{Soft X-ray properties of
  narrow-line Seyfert 1 galaxies.},'' {\em \aap}~{\bf 305},  53--+ (Jan. 1996).

\bibitem{2004MNRAS.349L...7G}
{Gierli{\'n}ski}, M. and {Done}, C., ``{Is the soft excess in active galactic
  nuclei real?},'' {\em \mnras}~{\bf 349},  L7--L11 (Mar. 2004).

\bibitem{2006MNRAS.365.1067C}
{Crummy}, J., {Fabian}, A.~C., {Gallo}, L., and {Ross}, R.~R., ``{An
  explanation for the soft X-ray excess in active galactic nuclei},'' {\em
  \mnras}~{\bf 365},  1067--1081 (Feb. 2006).

\bibitem{2015JHEAp...7..148K}
{Komossa}, S., ``{Tidal disruption of stars by supermassive black holes: Status
  of observations},'' {\em Journal of High Energy Astrophysics}~{\bf 7},
  148--157 (Sept. 2015).

\bibitem{2020MNRAS.498..859D}
{D'Ammando}, F., ``{Short time-scale variability of
  {\ensuremath{\gamma}}-ray-emitting narrow-line Seyfert 1 galaxies in optical
  and UV bands},'' {\em \mnras}~{\bf 498},  859--874 (Aug. 2020).

\bibitem{2018A&A...614A..87B}
{Berton}, M., {Congiu}, E., {J{\"a}rvel{\"a}}, E., {Antonucci}, R., {Kharb},
  P., {Lister}, M.~L., {Tarchi}, A., {Caccianiga}, A., {Chen}, S., {Foschini},
  L., {L{\"a}hteenm{\"a}ki}, A., {Richards}, J.~L., {Ciroi}, S., {Cracco}, V.,
  {Frezzato}, M., {La Mura}, G., and {Rafanelli}, P., ``{Radio-emitting
  narrow-line Seyfert 1 galaxies in the JVLA perspective},'' {\em \aap}~{\bf
  614},  A87 (June 2018).

\bibitem{2018ApJ...856....1P}
{Pasham}, D.~R. and {van Velzen}, S., ``{Discovery of a Time Lag between the
  Soft X-Ray and Radio Emission of the Tidal Disruption Flare ASASSN-14li:
  Evidence for Linear Disk-Jet Coupling},'' {\em \apj}~{\bf 856},  1 (Mar.
  2018).

\bibitem{2020NewAR..8901538D}
{De Colle}, F. and {Lu}, W., ``{Jets from Tidal Disruption Events},'' {\em New
  Astronomy Reviews}~{\bf 89},  101538 (Sept. 2020).

\bibitem{2015MNRAS.449..129W}
{Wilkins}, D.~R. and {Gallo}, L.~C., ``{Driving extreme variability: the
  evolving corona and evidence for jet launching in Markarian 335},'' {\em
  \mnras}~{\bf 449},  129--146 (May 2015).

\bibitem{2020arXiv200714704J}
{Jin}, C., {Done}, C., and {Ward}, M., ``{Re-observing the NLS1 Galaxy RE
  J1034+396. II. New Insights on the Soft X-ray Excess, QPO and the Analogy
  with GRS 1915+105},'' {\em arXiv e-prints} ,  arXiv:2007.14704 (July 2020).

\bibitem{weisskopf:77320E}
Weisskopf, M.~C., Elsner, R.~F., and O'Dell, S.~L., ``On understanding the
  figures of merit for detection and measurement of x-ray polarization,'' in
  [{\em Society of Photo-Optical Instrumentation Engineers (SPIE) Conference
  Series}{\nolinebreak\hspace{0.1em}]},  Arnaud, M., Murray, S.~S., and
  Takahashi, T., eds., {\em Space Telescopes and Instrumentation 2010:
  Ultraviolet to Gamma Ray}~{\bf 7732},  77320E, SPIE (2010).

\bibitem{2018NatAs...2..652C}
{Chauvin}, M., {Flor{\'e}n}, H.-G., {Friis}, M., {Jackson}, M., {Kamae}, T.,
  {Kataoka}, J., {Kawano}, T., {Kiss}, M., {Mikhalev}, V., {Mizuno}, T.,
  {Ohashi}, N., {Stana}, T., {Tajima}, H., {Takahashi}, H., {Uchida}, N., and
  {Pearce}, M., ``{Accretion geometry of the black-hole binary Cygnus X-1 from
  X-ray polarimetry},'' {\em Nature Astronomy}~{\bf 2},  652--655 (June 2018).

\bibitem{redsoxjatis}
{Marshall}, H.~L., {G{\"u}nther}, H.~M., {Heilmann}, R.~K., {Schulz}, N.~S.,
  {Egan}, M., {Hellickson}, T., {Heine}, S.~N.~T., {Windt}, D.~L., {Gullikson},
  E.~M., {Ramsey}, B.~D., {Tagliaferri}, G., and {Pareschi}, G., ``{Design of a
  Broad-band Soft X-ray Polarimeter},'' {\em Journal of Astronomical
  Telescopes, Instruments, and Systems}~{\bf 4},  11004 (Mar. 2018).

\bibitem{2019JATIS...5b1012Z}
{Zhang}, W.~W., {Allgood}, K.~D., {Biskach}, M.~P., {Chan}, K.-W., {Hlinka},
  M., {Kearney}, J.~D., {Mazzarella}, J.~R., {McClelland}, R.~S., {Numata}, A.,
  {Riveros}, R.~E., {Saha}, T.~T., and {Solly}, P.~M., ``{High-resolution,
  lightweight, and low-cost x-ray optics for the Lynx observatory},'' {\em
  Journal of Astronomical Telescopes, Instruments, and Systems}~{\bf 5},
  021012 (Apr 2019).

\bibitem{ogel93}
{Oegelman}, H., {Finley}, J.~P., and {Zimmerman}, H.~U., ``{Pulsed X-rays from
  the VELA pulsar},'' {\em \nat}~{\bf 361},  136--138 (Jan. 1993).

\bibitem{1987ApOpt..26.2915C}
{Cash}, W., ``{X-ray optics - A technique for high resolution imaging},'' {\em
  \ao}~{\bf 26},  2915--2920 (July 1987).

\bibitem{2009SPIE.7437E..14H}
{Heilmann}, R.~K., {Ahn}, M., {Bautz}, M.~W., {Foster}, R., {Huenemoerder},
  D.~P., {Marshall}, H.~L., {Mukherjee}, P., {Schattenburg}, M.~L., {Schulz},
  N.~S., and {Smith}, M., ``{Development of a critical-angle transmission
  grating spectrometer for the International X-Ray Observatory},'' in [{\em
  Optics for EUV, X-Ray, and Gamma-Ray Astronomy
  IV}{\nolinebreak\hspace{0.1em}]},  {\em \procspie} {\bf 7437},  74370G (Aug.
  2009).

\bibitem{2015SPIE.9603E..14H}
{Heilmann}, R.~K., {Bruccoleri}, A.~R., and {Schattenburg}, M.~L.,
  ``{High-efficiency blazed transmission gratings for high-resolution soft
  x-ray spectroscopy},'' in [{\em Society of Photo-Optical Instrumentation
  Engineers (SPIE) Conference Series}{\nolinebreak\hspace{0.1em}]},  {\em
  \procspie} {\bf 9603},  960314 (Sept. 2015).

\bibitem{heilmann17}
{Heilmann}, R.~K., {Bruccoleri}, A.~R., {Song}, J., {Kolodziejczak}, J.,
  {Gaskin}, J.~A., {O'Dell}, S.~L., {Cheimetz}, P., {Hertz}, E., {Smith},
  R.~K., {Burwitz}, V., {Hartner}, G., {La Caria}, M.-M., and {Schattenburg},
  M.~L., ``{Critical-angle transmission grating technology development for high
  resolving power soft x-ray spectrometers on Arcus and Lynx},'' in [{\em
  Society of Photo-Optical Instrumentation Engineers (SPIE) Conference
  Series}{\nolinebreak\hspace{0.1em}]},  {\em Society of Photo-Optical
  Instrumentation Engineers (SPIE) Conference Series} {\bf 10399},  1039914
  (Aug. 2017).

\bibitem{2017SPIE10397E..0QS}
{Smith}, R.~K., {Abraham}, M., {Allured}, R., {Bautz}, M., {Bookbinder}, J.,
  {Bregman}, J., {Brenneman}, L., {Brickhouse}, N.~S., {Burrows}, D.,
  {Burwitz}, V., {Cheimets}, P.~N., {Costantini}, E., {Dawson}, S., {DeRoo},
  C., {Falcone}, A., {Foster}, A.~R., {Gallo}, L., {Grant}, C.~E.,
  {G{\"u}nther}, H.~M., {Heilmann}, R.~K., {Hertz}, E., {Hine}, B.,
  {Huenemoerder}, D., {Kaastra}, J.~S., {Kreykenbohm}, I., {Madsen}, K.~K.,
  {McEntaffer}, R., {Miller}, E., {Miller}, J., {Morse}, E., {Mushotzky}, R.,
  {Nandra}, K., {Nowak}, M., {Paerels}, F., {Petre}, R., {Poppenhaeger}, K.,
  {Ptak}, A., {Reid}, P., {Sanders}, J., {Schattenburg}, M., {Schulz}, N.,
  {Smale}, A., {Temi}, P., {Valencic}, L., {Walker}, S., {Willingale}, R.,
  {Wilms}, J., and {Wolk}, S.~J., ``{Arcus: exploring the formation and
  evolution of clusters, galaxies, and stars},'' in [{\em Society of
  Photo-Optical Instrumentation Engineers (SPIE) Conference
  Series}{\nolinebreak\hspace{0.1em}]},  {\em Society of Photo-Optical
  Instrumentation Engineers (SPIE) Conference Series} {\bf 10397},  103970Q
  (Aug. 2017).

\bibitem{2017SPIE10399E..15S}
{Song}, J., {Heilmann}, R.~K., {Bruccoleri}, A.~R., {Hertz}, E., and
  {Schatternburg}, M.~L., ``{Scanning laser reflection tool for alignment and
  period measurement of critical-angle transmission gratings},'' in [{\em
  Society of Photo-Optical Instrumentation Engineers (SPIE) Conference
  Series}{\nolinebreak\hspace{0.1em}]},  {\em Society of Photo-Optical
  Instrumentation Engineers (SPIE) Conference Series} {\bf 10399},  1039915
  (Aug. 2017).

\bibitem{2014SPIE.9144E..1KM}
{Marshall}, H.~L., {Schulz}, N.~S., {Windt}, D.~L., {Gullikson}, E.~M.,
  {Blake}, E., {Getty}, D., and {McInturff}, Z., ``{The use of laterally graded
  multilayer mirrors for soft X-ray polarimetry},'' in [{\em Society of
  Photo-Optical Instrumentation Engineers (SPIE) Conference
  Series}{\nolinebreak\hspace{0.1em}]},  {\em \procspie} {\bf 9144},  1 (July
  2014).

\bibitem{2015SPIE.9603E..19M}
{Marshall}, H.~L., {Schulz}, N.~S., {Windt}, D.~L., {Gullikson}, E.~M.,
  {Craft}, M., {Blake}, E., and {Ross}, C., ``{The use of laterally graded
  multilayer mirrors for soft x-ray polarimetry},'' in [{\em Society of
  Photo-Optical Instrumentation Engineers (SPIE) Conference
  Series}{\nolinebreak\hspace{0.1em}]},  {\em \procspie} {\bf 9603},  960319
  (Sept. 2015).

\bibitem{2008SPIE.7011E..2CL}
{LaMarr}, B., {Grant}, C., {Kissel}, S., {Prigozhin}, G., {Bautz}, M., {Tsuru},
  T.~G., {Tsunemi}, H., {Dotani}, T., {Hayashida}, K., and {Matsumoto}, H.,
  ``{Front- and back-illuminated x-ray CCD performance in low- and high-Earth
  orbit: performance trends of Suzaku XIS and Chandra ACIS detectors},'' in
  [{\em Space Telescopes and Instrumentation 2008: Ultraviolet to Gamma
  Ray}{\nolinebreak\hspace{0.1em}]},  {\em \procspie} {\bf 7011},  70112C (July
  2008).

\bibitem{2017AJ....154..243G}
{G{\"u}nther}, H.~M., {Frost}, J., and {Theriault-Shay}, A., ``{MARXS: A
  Modular Software to Ray-trace X-Ray Instrumentation},'' {\em \aj}~{\bf 154},
  243 (Dec. 2017).

\bibitem{2009PASJ...61.1117B}
{Bautz}, M.~W., {Miller}, E.~D., {Sanders}, J.~S., {Arnaud}, K.~A.,
  {Mushotzky}, R.~F., {Porter}, F.~S., {Hayashida}, K., {Henry}, J.~P.,
  {Hughes}, J.~P., {Kawaharada}, M., {Makashima}, K., {Sato}, M., and {Tamura},
  T., ``{Suzaku Observations of Abell 1795: Cluster Emission to r$_{200}$},''
  {\em \pasj}~{\bf 61},  1117 (Oct. 2009).

\bibitem{murphy:77322Y}
{Murphy}, K.~D., {Marshall}, H.~L., {Schulz}, N.~S., {Jenks}, K., {Sommer}, S.
  J.~B., and {Marshall}, E.~A., ``{Soft x-ray polarimeter laboratory tests},''
  in [{\em Space Telescopes and Instrumentation 2010: Ultraviolet to Gamma
  Ray}{\nolinebreak\hspace{0.1em}]},  {\em Society of Photo-Optical
  Instrumentation Engineers (SPIE) Conference Series} {\bf 7732},  77322Y (July
  2010).

\bibitem{2013SPIE.8861E..1DM}
{Marshall}, H.~L., {Schulz}, N.~S., {Remlinger}, B., {Gentry}, E.~S., {Windt},
  D.~L., and {Gullikson}, E.~M., ``{Progress toward a soft x-ray
  polarimeter},'' in [{\em Society of Photo-Optical Instrumentation Engineers
  (SPIE) Conference Series}{\nolinebreak\hspace{0.1em}]},  {\em \procspie} {\bf
  8861} (Sept. 2013).

\bibitem{garner19}
{Garner}, A., {Marshall}, H., {Heine}, S., {Heilmann}, R., {Song}, J.,
  {Schulz}, N., {LaMarr}, B., and {Egan}, M., ``{Component testing for x-ray
  spectroscopy and polarimetry},'' in [{\em Society of Photo-Optical
  Instrumentation Engineers (SPIE) Conference
  Series}{\nolinebreak\hspace{0.1em}]},  {\em \procspie} {\bf 11118},
  1111811--1--12 (Sept. 2019).

\bibitem{lynx}
{Schwartz}, D.~A., {Vikhlinin}, A., {Tananbaum}, H., {Freeman}, M., {Tremblay},
  G., {Schwartz}, E.~D., {Gaskin}, J.~A., {Swartz}, D., {Gelmis}, K.,
  {McCarley}, K.~S., and {Dominguez}, A., ``{The Lynx X-ray Observatory:
  revealing the invisible universe},'' in [{\em UV, X-Ray, and Gamma-Ray Space
  Instrumentation for Astronomy XXI}{\nolinebreak\hspace{0.1em}]},  {\em
  Society of Photo-Optical Instrumentation Engineers (SPIE) Conference Series}
  {\bf 11118},  111180K (Sept. 2019).

\bibitem{2019SPIE11118E..0JB}
{Bautz}, M.~W., ``{The Lynx X-Ray Observatory: Science Drivers},'' in [{\em UV,
  X-Ray, and Gamma-Ray Space Instrumentation for Astronomy
  XXI}{\nolinebreak\hspace{0.1em}]},  {\em Society of Photo-Optical
  Instrumentation Engineers (SPIE) Conference Series} {\bf 11118},  111180J
  (Sept. 2019).

\bibitem{xpp}
{Jahoda}, K., {Krawczynski}, H., {Kislat}, F., {Marshall}, H., {Okajima}, T.,
  {Agudo}, I., {Angelini}, L., {Bachetti}, M., {Baldini}, L., {Baring}, M.,
  {Baumgartner}, W., {Bellazzini}, R., {Bianchi}, S., {Bucciantini}, N.,
  {Caiazzo}, I., {Capitanio}, F., {Coppi}, P., {Costa}, E., {De Rosa}, A., {Del
  Monte}, E., {Dexter}, J., {Di Gesu}, L., {Di Lalla}, N., {Doroshenko}, V.,
  {Dovciak}, M., {Ferrazzoli}, R., {Fuerst}, F., {Garner}, A., {Ghosh}, P.,
  {Gonzalez-Caniulef}, D., {Grinberg}, V., {Gunji}, S., {Hartman}, D.,
  {Hayashida}, K., {Heyl}, J., {Hill}, J., {Ingram}, A., {Buz Iwakiri}, W.,
  {Jorstad}, S., {Kaaret}, P., {Kallman}, T., {Karas}, V., {Khabibullin}, I.,
  {Kitaguchi}, T., {Kolodziejczak}, J., {Kouveliotou}, C., {Liodakis}, Y.,
  {Maccarone}, T., {Manfreda}, A., {Marin}, F., {Marinucci}, A., {Markwardt},
  C., {Marscher}, A., {Matt}, G., {McConnell}, M., {Miller}, J., {Mitsubishi},
  I., {Mizuno}, T., {Mushtukov}, A., {Ng}, S., {Nowak}, M., {O'Dell}, S.,
  {Papitto}, A., {Pasham}, D., {Pearce}, M., {Peirson}, L., {Perri}, M., {Pesce
  Rollins}, M., {Petrosian}, V., {Petrucci}, P.-O., {Pilia}, M., {Possenti},
  A., {Poutanen}, J., {Prescod-Weinstein}, C., {Puccetti}, S., {Salmi}, T.,
  {Shi}, K., {Soffita}, P., {Spandre}, G., {Steiner}, J., {Strohmayer}, T.,
  {Suleimanov}, V., {Svoboda}, J., {Swank}, J., {Tamagawa}, T., {Takahashi},
  H., {Taverna}, R., {Tomsick}, J., {Trois}, A., {Tsygankov}, S., {Turolla},
  R., {Vink}, J., {Wilms}, J., {Wu}, K., {Xie}, F., {Younes}, G., {Zaino}, A.,
  {Zajczyk}, A., {Zane}, S., {Zdziarski}, A., {Zhang}, H., {Zhang}, W., and
  {Zhou}, P., ``{The X-ray Polarization Probe mission concept},'' {\em arXiv
  e-prints} ,  arXiv:1907.10190 (Jul 2019).

\bibitem{10.1117/1.JATIS.4.1.011004}
Kislat, F., Abarr, Q., Beheshtipour, B., Geronimo, G.~D., Dowkontt, P., Tang,
  J., and Krawczynski, H., ``{Optimization of the design of X-Calibur for a
  long-duration balloon flight and results from a one-day test flight},'' {\em
  Journal of Astronomical Telescopes, Instruments, and Systems}~{\bf 4}(1),  1
  -- 9 (2018).

\bibitem{2019arXiv190409313K}
{Krawczynski}, H., {Matt}, G., {Ingram}, A.~R., {Taverna}, R., {Turolla}, R.,
  {Kislat}, F., {Teddy Cheung}, C.~C., {Bykov}, A., {Sinha}, K., {Zhang}, H.,
  {Heyl}, J., {Bucciantini}, N., {Madejski}, G., {Kallman}, T., {Jahoda},
  K.~M., {Abarr}, Q., {Baring}, M.~G., {Baldini}, L., {Begelman}, M.,
  {Boettcher}, M., {Cackett}, E., {Caiazzo}, I., {Coppi}, P., {Costa}, E.,
  {Dexter}, J., {Fortin}, J.-F., {Gammie}, C., {Gaskin}, J.~A., {Giannios}, D.,
  {Ghosh}, P., {Harding}, A.~K., {Hartmann}, D.~H., {Hayashida}, K., {Jorstad},
  S.~G., {Kaaret}, P., {Kitaguchi}, T., {Latronico}, L., {Maccarone}, T.,
  {Marscher}, A., {Marshall}, H., {McConnell}, M., {Miller}, J.~M., {O'Dell},
  S.~L., {Oezel}, F., {Okajima}, T., {Pearce}, M., {Perkins}, J., {Ramsey},
  B.~D., {Romani}, R.~W., {Schnittman}, J.~D., {Sgro}, C., {Soffitta}, P.,
  {Takahashi}, H., {Tamagawa}, T., {Tomsick}, J., {Weisskopf}, M.~C., and
  {Younes}, G., ``{Astro2020 Science White Paper: Using X-Ray Polarimetry to
  Probe the Physics of Black Holes and Neutron Stars},'' {\em arXiv e-prints} ,
   arXiv:1904.09313 (Apr 2019).

\end{thebibliography}

\end{document}